\documentclass[]{interact}

\usepackage{epstopdf}
\usepackage[caption=false]{subfig}
\usepackage{xcolor}

\usepackage[numbers,sort&compress]{natbib}
\bibpunct[, ]{[}{]}{,}{n}{,}{,}
\makeatletter
\def\NAT@def@citea{\def\@citea{\NAT@separator}}
\makeatother

\theoremstyle{plain}

\theoremstyle{definition}

\theoremstyle{remark}

\numberwithin{equation}{section}

\begin{document}


\title{ELECTRON TRANSPORT AND ELECTRON DENSITY INSIDE ONE-DIMENSIONAL 
DISORDERED CONDUCTORS: \\
An Analysis of the Electronic-Levels Contribution}

\author{
\name{Gerardo Rivas\textsuperscript{a}
\thanks{},
Miztli Y\'epez\textsuperscript{b}
\thanks{CONTACT M. Y\'epez. Email: miztli.yolotzin.yepez@uacm.edu.mx}
and
Pier A. Mello\textsuperscript{a}}
\affil{\textsuperscript{a}Instituto de F\'isica, Universidad Nacional Aut\'onoma de M\'exico 
Ciudad de M\'exico, 04510, Mexico.
\\
\textsuperscript{b}Academia de F\'isica, Universidad Aut\'onoma de la Ciudad de M\'exico-Casa Libertad. Ciudad de M\'exico, 09620, Mexico.}
}


\maketitle

\begin{abstract}
We consider the problem of electron transport along a one-dimensional disordered multiple-scattering conductor, and study the electron density for all the electronic levels. A model is proposed for the reduced density matrix of the system placed between two reservoirs at different chemical potentials, and the statistical-mechanical expectation value of the electron density is evaluated. An ensemble average is computed over disordered configurations. We compare its predictions with computer simulations. We find that the contribution of low-lying levels is very different from that of the high-lying ones studied in the past. Going down in energy, the wave function penetrates ever less inside the sample. For high-lying levels, this is interpreted in terms of localization from disorder. For low-lying levels, this interpretation gradually gives way to an understanding in terms of the increasing reflection produced by each scatterer, which is 'seen' by the electron as a higher and higher --and hence impenetrable-- potential barrier. Indeed, the local-density-of-states, LDOS, is gradually depleted in the interior of the system, since the wave function is ever smaller inside. The problem studied here is also of interest in electromagnetic, thermal, and acoustic transport in disordered systems.
\end{abstract}


\begin{pacscode}
73.23.-b, 05.60.Gg, 72.10.-d, 11.55.-m
\end{pacscode}



\section{Introduction}
\label{intro}

The physics of electronic transport in disordered mesoscopic systems has been studied for many years (for a review, see Ref.  \cite{mello-kumar} and references cited therein). The discovery by Landauer \cite{landauer} of the electronic conductance being proportional to the transmittance was a breakthrough in the investigation of mesoscopic systems, since it has allowed studying the conductance in terms of the scattering properties of the system under investigation \cite{landauer,buettiker,mello-kumar}. This equivalence allows many of the predictions of mesoscopic physics and localization theory to apply to the transport of quantum as well as classical waves  \cite{Thouless_1977,Abrahams_1979,Anderson_1980,Melnikov_1980,Dorokhov_1984,van_Rossum_1999,
Shi_2014,Dietz_2011,dahr}.

The conductance, the transmittance and their statistical properties refer to physical quantities evaluated {\it outside} the sample. Besides these properties, the problem of the statistics of transport {\it inside} random systems has also been studied for many years \cite{gazaryan,kohler_papanicolau,genack1990long,Neupane-Yamilov,van_tiggelen_et_al_2000-2006,van_tiggelen_et_al_2000-2006_bis,tian_et_al_2010,mello-shi-genack,cheng-yepez-mello-genack}; 
see also Ref. \cite{boris_1986}, in which the author studied the short-range intensity correlations with both the source and the detector being inside the (infinite) medium, and Ref.  \cite{boris_2014}, where the authors measured the long-range intensity correlation inside the sample, placed  between two leads. 
In a series of recent papers, the control of different mesoscopic transport effects inside random media has been studied by changing the system's geometry, or shaping incident wavefronts \cite{boris_2014,bender_coherent_2022,bender_depth-targeted_2022,
bender_introducing_2019,koirala_inverse_2019,sarma_control_2015,sarma_using_2015,
sarma_control_2016,yamilov_shape_2016,yamilov_sum_2022,
yilmaz_2019_transverse}.

The problem is of interest in various branches of physics, the reason being that it is representative of a rather general wave-scattering problem: e.g., an electromagnetic wave traveling in a disordered waveguide \cite{cheng-yepez-mello-genack} --the interest then being in the energy density inside the structure--, or an elastic wave propagating in a disordered elastic waveguide \cite{elastic_waves,elastic_waves_2} --the interest then being, e.g., the mean square displacement inside the system.

One-dimensional (1D) disordered systems were studied very intensely in the 1980's and 1990's, as they can be considered as the simplest realizations of such problems. Some representative contributions are given, e.g., by Refs. \cite{Anderson_1980,Erdos,Shapiro_1986,mello_1987}.

Interestingly enough, the density (for particles) or intensity (for waves) inside the sample for 1D systems are still of interest in the present time, the motivation being the following.

\begin{enumerate}

\item [1] There is great interest in experiments with cold atoms (matter waves) in 1D channels. 
The atoms can be either bosons or fermions- the latter case should be similar to electrons.
The reader is referred, e.g., to Refs.  \cite{Aspect_2009,Jean-Philippe} and Refs.  \cite{Billy_et_al,Roati_et_al}: the last two are experiments concerning localization of cold atoms in 1D.

\item [2] For electromagnetic waves, of course one can probe the intensity inside. 
Two concrete examples are the recent experiments reported in Refs. \cite{cheng-yepez-mello-genack, Huang_2022}, in which waves are launched from one end of the waveguide and the signal is detected by an antenna just above a slit along the length of the waveguide.

\item [3] There is a renewed interest in 1D disordered problems, in connection with
'temporarily modulated media' (see, e.g., Ref.  \cite{Carminati}). 
In that problem, space is homogeneous,
but the dielectric function $\epsilon(t)$ is a random function of time. 
One considers an initial wave packet 
with a well defined wave vector,
and asks how this wave behaves in time.

\item [4] The dynamic approach to quantum transport has been studied since the early investigations \cite{buttiker_larmor_nodate}, and recent advancements have been done in this respect, particularly in the study of the transmission eigenchannels \cite{winful_delay_2003, iannaccone_general_1994, shi2015dynamic, choi, davy2015transmission}.

\item [5] Finally, in the last decade there have been advancements in the understanding of the fluctuations and correlations of scattering properties inside the sample \cite{genack1990long, shi2015dynamic, bender2020fluctuations}.

\end{enumerate}

Formally, the problem is equivalent to a 1D random chain, similar to 1D Schr\"odinger's equation (but the second derivative is in time rather than in coordinate).  

Similarly, in our group we have recently studied the statistical properties of the electron density inside a multiply scattering 1D disordered system \cite{mello-shi-genack,cheng-yepez-mello-genack}, 
and also its extension to a quasi-one-dimensional (q1D)  \cite{mello-yepez} disordered geometry. 
In those studies, the system was fed with electrons of a {\it given energy} from one end of the disordered conductor and the electron density was evaluated along the conductor and outside. In Ref.  \cite{mello-shi-genack}, 
the expectation value $\langle {\cal{W}}(x)\rangle$ of the intensity ${\cal W}(x)$ at a distance $x$ from the entrance 
of a 1D disordered system
was calculated and compared with computer simulations. In Ref. \cite{cheng-yepez-mello-genack}, also for 1D systems, emphasis was put on the statistics of the logarithm  of the intensity, $\ln {\cal{W}}(x)$, which shows interesting scaling properties, in a way similar to the logarithm of the conductance in the conduction problem; theoretical predictions were compared with computer simulations and also with the results of microwave experiments. More recently  \cite{mello-yepez}, we studied the statistical properties of the electron density inside a q1D multiply-scattering medium, i.e., a system supporting more than one propagating mode or open channel.

We should remark that in a real electrical conduction problem realized by inserting the system between two terminals (reservoirs) at different chemical potentials, the electron density inside the system would have to be calculated by adding the contribution of {\it all incident energies} at which electrons are fed by the reservoirs, with a weight given by the Fermi function of the respective reservoir. Thus, whereas in the above-mentioned papers the analysis was restricted to one energy, the more complete calculation is the purpose of the present paper. This study may encourage the development of methods toward the experimental verification of its results. One possible experimental setup was described in Ref.  \cite{mello-yepez} for the case of one energy.

In the whole paper, our goal will be to study a 1D electronic conductor in a scheme of non-interacting electrons moving in a self-consistent potential.

We first set up the statistical mechanical problem for a given sample, and then consider a collection of samples in order to construct an ensemble of configurations of disorder, both theoretically and through computer simulations. The specific theoretical model that our computer simulations will be compared with is designated as the DMPK model \cite{mello-kumar,Dorokhov_1984,mpk}; for 1D disordered system, the DMPK equation reduces to Melnikov's \cite{Melnikov_1980}.

The present analysis extends and generalizes the results  for 1D systems of Refs.  \cite{mello-shi-genack,cheng-yepez-mello-genack} which dealt with high-lying levels only, in order to take into account low-lying levels. The behavior of the latter is found to be very different from that of the high-lying ones: this is illustrated in Fig. \ref{density for L and for R incid aa} below. 
This constitutes the main difference with respect to our previous work; it is discussed and well understood and represents one of our main results.  
While DMPK gives a good description of the high-lying levels, for the low-lying ones it does not;
for these latter states we have, at present, no theoretical model.

The paper is organized as follows. In the next section, we start our discussion with a ballistic, non-disordered, mesoscopic conductor. We then discuss generalities of a disordered 1D system, which are subsequently applied to the  study of the electron density, which is our main interest in the present paper. In Sec. \ref{analysis electron density}, we find explicit expressions for the electron density in the various regions --inside the disordered sample and outside-- and construct the corresponding expectation values over an ensemble of configurations of disorder. In Sec. \ref{expect_density_dmpk_simul}, we study the electron density for a system in equilibrium at zero temperature and its expectation value over disorder. We analyze the contribution of single levels, and then the result obtained taking all levels into account. In Sec. \ref{aver log density}, we extend the analysis to the logarithm of the electron density. In Sec. \ref{density_dmpk_no_equil}, we extend the above results to a non-equilibrium --but stationary-- situation, in which the chemical potentials of the two reservoirs are not equal. Finally, we present our conclusions in Sec. \ref{conclusions}. Some technical details can be found in the appendices.

Various appendices are included to prove certain specific results without interrupting the main flow of the paper.
\section{The electron density inside a 1D conductor}
\label{density 1D}

In the whole paper our goal will be to study a one-dimensional electronic conductor placed between two reservoirs at temperature $T$ and chemical potentials $\mu_1$ and $\mu_2$. 
When $\mu_1=\mu_2$, the system of interest is in thermodynamic equilibrium. We shall assume an approximate picture: in equilibrium, the system is described by a Hamiltonian of non-interacting electrons moving in a self-consistent potential: each electron interacts with the average electron density of the rest of electrons rather than with each electron individually; we also include confining potentials, and ions and impurities with no internal degrees of freedom, thus producing only elastic scattering 
(see Refs.  \cite{mello-kumar,mello-shapiro-imry} and references contained therein).
For simplicity, however, in this section we first discuss
the case of a ballistic conductor, in order to pave the way to the general analysis of an arbitrary disordered conductor.

\subsection{Ballistic 1D conductor: generalities}
\label{ballistic 1D}

As a preliminary study, the system to be considered in this section will be a ballistic 1D electronic conductor of length $L_0$. For simplicity, we shall ignore spin-orbit coupling and deal with 'spinless electrons' \cite{mello-kumar}. We define, in the interior of the conductor, a complete set of orthonormal states with periodic boundary conditions, which represent running waves:
\begin{subequations}
\begin{eqnarray}
\phi_{n}(x) &=& \frac{e^{i s_n k_n x}}{\sqrt{L_0}} 
\;\;\;\;\;
\left\{
\begin{array}{c}
n= 0,\pm 1, \pm 2, \cdots  \\
s_n = {\rm sgn}(n)=\pm 1
\end{array}
\right. \; ,
\label{phi s}  \\
k_n &=&\frac{2\pi | n |}{L_0},   \,
\label{ks}   \\
\epsilon_n &=& \frac{\hbar^2 k_n^2}{2m} \; .
\label{Es}
\end{eqnarray}
\label{PBC states}
\end{subequations}
The wavenumber $k_n$ is defined to be {\it positive}, so the direction of propagation is specified by $s_n = \pm 1$. The total number of states 
${\cal N}^{+}(\epsilon_n)$ 
traveling to the right, up to the energy $\epsilon_n$, and the corresponding density of states $\rho^{+}(\epsilon)$, i.e, the number of states per unit energy around the energy $\epsilon$ (a continuous function of $\epsilon$), are given, respectively, by
\begin{subequations}
\begin{eqnarray}
{\cal N}^{+}(\epsilon_n)
&=& \frac{k_nL_0}{2 \pi},
\label{number of states} \\
\rho^{+}(\epsilon) 
&=&\frac{\partial {\cal N}^{+}(\epsilon)}{\partial \epsilon}
= \frac{L_0}{2 \pi \hbar}\frac{1}{v_{gr}}, \;\;\;  
\label{density of states b} \\
v_{gr}&=& \frac{\hbar k}{m} 
= \sqrt{\frac{2\epsilon}{m}},
\label{density of states c}
\end{eqnarray}
\label{number and density of states}
\end{subequations}
where $v_{gr}$ designates the group velocity. We have similar expressions for electrons traveling to the left.

\begin{figure}[b]
\centerline{
\includegraphics[width=\columnwidth]{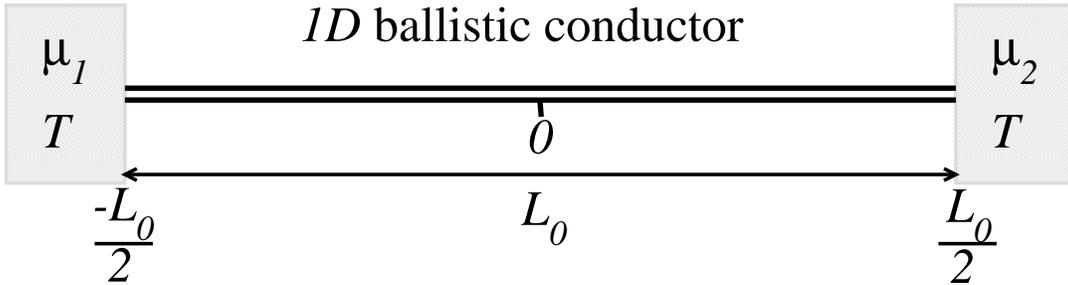}
}
\caption{
Schematic representation of a ballistic 1D electronic conductor of length $L_0$ placed between two reservoirs at chemical potentials $\mu_1 > \mu_2$. 
}
\label{ballistic 1D and two reservoirs}
\end{figure}

We follow Ref.  \cite{buettiker} and describe our system of interest as  illustrated in Fig. \ref{ballistic 1D and two reservoirs}: on the left, it is in contact with reservoir 1, which has temperature $T$ and chemical potential $\mu_1$; on the right, it is in contact with reservoir 2, at the same temperature $T$, and chemical potential $\mu_2$. The reservoirs emit electrons toward the system and absorb, without reflection, the electrons incident upon them.

We do not wish to include the reservoirs in our description, but only the 'system proper'. We thus trace the complete density matrix over the reservoirs degrees of freedom: the result is what is called the {\it reduced density matrix}.

When $\mu_{1} \neq \mu_{2}$, the system is not in equilibrium, but we suppose it is in a stationary state.
The density matrix $\widehat{\rho}(\beta, \mu_1, \mu_2)$ describes a {\it non-equilibrium}, although {\it stationary state}.
We introduce the following simple model to describe the system reduced density matrix. Let $c_{n}^{\dagger}$, $c_{n}$ be the electron creation and annihilation operators
associated with the single-electron state $\phi_n(x)$ of Eq. (\ref{phi s}) with energy $\epsilon_{n}$ ($\epsilon_{0}=0$), and let $\mu_{0}\equiv(\mu_{1} + \mu_{2})/2$. Then the model is defined as
\begin{eqnarray}
\widehat{\rho}(\beta, \mu_1, \mu_2)
&=&  
\frac{  e^{-\beta \sum_{n > 0} (\epsilon_n - \mu_1) c^{\dagger}_n c_n  }}
{{\cal Z}^{(+)}(\beta, \mu_1)} 
\times 
\frac{  e^{-\beta (\epsilon_0 - \mu_0) c^{\dagger}_0 c_0  }}
{1+ e^{-\beta(\epsilon_0 - \mu_0)}} 
\times
\frac{  e^{-\beta \sum_{n < 0} (\epsilon_n - \mu_2) c^{\dagger}_n c_n  }}
{{\cal Z}^{(-)}(\beta, \mu_2)},
\label{model density matrix a} 
\nonumber
\\  
\end{eqnarray}
which reduces to the equilibrium grand canonical density matrix when the chemical potentials are equal, $\mu_1=\mu_0=\mu_2$. For a more formal analysis of the problem using linear-response theory, the reader may consult, e.g., Refs.  \cite{mello-kumar,mello-shapiro-imry} and references cited therein.

It can be shown that the arguments given by B\"uttiker,  \cite{buettiker} leading to the well-known expression for the current through a 1D conductor and the associated conductance for the present case, are equivalent to using the above simple model (although the reduced density matrix is not mentioned explicitly in Ref.  \cite{buettiker}).

Thus the current through a 1D ballistic conductor and the associated conductance can be computed by using the above model for the reduced density matrix to find B\"uttiker’s well-known result for this particular case  \cite{buettiker}. A similar procedure allows us to find the electron density in a 1D ballistic conductor.

The partition functions are given by
\begin{subequations}
\begin{eqnarray}
{\cal Z}^{(+)}(\beta, \mu_1)
&=& \prod_{n > 0} \left[1 + e^{-\beta(\epsilon_n - \mu_1)}     \right] ,
\label{model density matrix c} \\
{\cal Z}^{(0)}(\beta, \mu_0) ,
&=& 1+ e^{\beta \mu_0} ,  \\
{\cal Z}^{(-)}(\beta, \mu_2)
&=& \prod_{n < 0} \left[1 + e^{-\beta(\epsilon_n - \mu_2)}     \right]  .
\label{model density matrix d}
\end{eqnarray}
\label{model density matrix}
\end{subequations}

We also mention Ref.  \cite{dhar_saito_hanggi}, which presents a non-equilibrium density matrix description of steady-state quantum transport. In App. \ref{dhar_et_al_density_matrix} we give the correspondence between the above ansatz (\ref{model density matrix}) and the results given in Ref.  \cite{dhar_saito_hanggi}.

\subsection{Disordered 1D conductor: generalities} 
\label{disordered}

The actual system of interest in this paper, the 'system proper' (disordered system), consists of independent electrons interacting with $N_{scatt}$ scattering units, numbered $j=1, \cdots, N_{scatt}$, sampled from some statistical distribution to be specified later, and occupying a length $L$. The single-particle  Hamiltonian thus consists of this single-particle potential plus the single-particle kinetic energy. When we speak of the full system, we mean the disordered system plus the leads.	

We solve this problem in two steps, as we now explain.

\subsubsection{The single-particle scattering problem in the interval $-\infty < x < +\infty$}

\begin{figure}[h]
\centerline{
\includegraphics[width=\columnwidth]{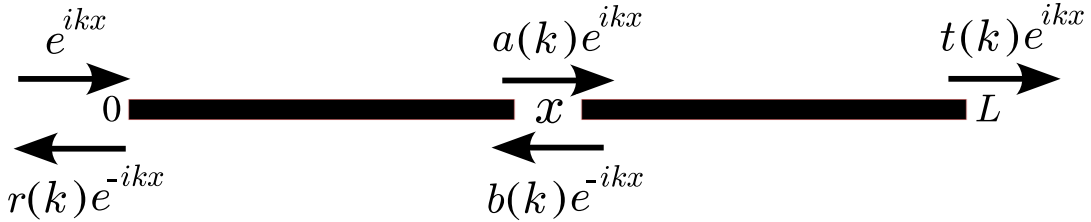}
}
\caption{
Schematic representation of the scattering problem for $\psi_{+, k}(x)$ of Table \ref{structure of wf}.
}
\label{Squematich_amplitudes_1}
\end{figure}

We first solve the single-particle scattering problem in the interval $-\infty < x < +\infty$ for the full system, consisting of the sample to which we have added perfect conductors on both sides, as illustrated in Fig. \ref{disordered 1D in (-infty, + infty)}. We designate by $\psi_{s, k}(x)$ the resulting eigenfunctions. They are shown in Table \ref{structure of wf}, where $r(k)$, $t(k)$, $r'(k)$, $t'(k)$ denote reflection and transmission amplitudes, and $a(k)$, $b(k)$, $a'(k)$, $b'(k)$ the appropriate amplitudes between two successive scattering units inside the disordered region;
we are assuming that between individual scatterers there is a free potential region, with a `small' width, where the wave function can be written as it is indicated in the second column of Table \ref{structure of wf}: see Fig. \ref{Squematich_amplitudes_1}

The wave functions $\psi_{s, k}(x)$ form a {\it complete set of orthonormal states}
 in the interval 
$x \in (-\infty, \infty)$.
\begin{figure}[t]
\centerline{
\includegraphics[width=\columnwidth]{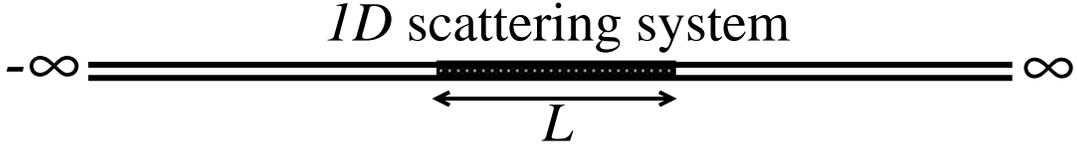}
}
\caption{
Schematic representation of a 1D scattering system of length $L$, with 'clean' regions extending to $=-\infty$ on the left and to $=+\infty$ on the right.
}
\label{disordered 1D in (-infty, + infty)}
\end{figure}
\begin{table}[h]
\tbl{The structure of the wave function $\psi_{s,k}(x)$ in the three regions shown in Fig. \ref{disordered 1D in (-infty, + infty)}, when incidence is from the left, $s=+$ (2nd row), and when incidence is from the right, $s=-$ (3rd row). The indicated wave function inside the sample is that occurring between two successive scattering units.}
{
\begin{tabular}{| c | c | c | c | 
}
\hline
& $-\infty < x \leq 0$
& $0 \leq x \leq L$ & $L \leq x < \infty$
\\
\hline\hline
$\psi_{+, k}(x)$
&
$ \frac{1}{\sqrt{2 \pi}}
\left[  e^{ik x} + r(k) e^{-ik x}
\right] $
& $ \frac{1}{\sqrt{2 \pi}}
\left[  a(k)  e^{ik x} + b(k)  e^{-ik x} \right] $ 
& $\; \frac{1}{\sqrt{2 \pi}} t(k) e^{ik x} $
\\
\hline
$  \psi_{-, k}(x)      $
& $ \frac{1}{\sqrt{2 \pi}} t'(k) e^{-ik x} $
& $ \frac{1}{\sqrt{2 \pi}}
\left[a'(k)  e^{ik x} + b'(k)  e^{-ik x} \right] $ & 
$\frac{1}{\sqrt{2 \pi}}
\left[ e^{-ik x}  + r'(k)e^{ik x} \right] $
\\
\hline\hline
\end{tabular}}
\label{structure of wf}
\end{table}


\subsubsection{The single-particle scattering problem in the interval $-L_{0}/2 < x < L_{0}/2$}

We restrict our full system to the interval $x\in\left(-L_{0}/2,L_{0}/2 \right)$, with $L_0 > L$. The functions $\psi_{s,k}(x)$ of Table \ref{structure of wf}, restricted to that interval, i.e.,
\begin{subequations}
\begin{equation}
\psi_{s,k}(x) \cdot \theta_{L_0}(x), 
\label{Restricted_WaveFunctions}
\end{equation}
where
\begin{eqnarray}
\theta_{L_0}(x) 
&=&
\left\{
\begin{array}{cc}
1, & x\in(-L_0/2, L_0/2)
\\
0, & x\notin(-L_0/2, L_0/2)
\end{array}
\right. \; ,
\end{eqnarray}
\end{subequations}
satisfy Schr\"odinger's equation in that interval, 
but form an {\it over-complete} set of states.

If we now consider the {\it subset} specified by the wavenumbers $k=k_n= 2\pi |n|/L_0$ of Eq. \eqref{ks}, the functions
\begin{subequations}
\begin{equation}
\psi_{s_n,k_n}^{L_0}(x) 
= \sqrt{ \frac{2\pi}{L_{0}} } \psi_{s_{n},k_n}(x) \cdot \theta_{L_0}(x)
=\left[ \phi_{n}\left(x\right) + \psi_{s_{n},k_n}^{scatt}(x) \right] \cdot \theta_{L_0}(x)
\label{cut-down wfs}
\end{equation}
satisfy Schr\"odinger's equation in $x\in (-L_0/2, L_0/2)$, i.e.,
\begin{eqnarray}
H  \psi_{s_{n},k_n}^{L_0}(x)  
&=& \epsilon_n \psi_{s_{n},k_n}^{L_0}(x), \;\;\;\; 
\end{eqnarray}
\end{subequations}
and consist of a {\it complete set of orthonormal unperturbed states} $\phi_{n}\left(x\right)$, Eq. \eqref{phi s}, plus scattered states $\psi_{s_{n},k_n}^{scatt}(x)$. They will be designated as $\psi_{n}^{L_0}(x) \equiv \psi_{s_{n}, k_n}^{L_0}(x)$. We have analytical and numerical evidence 
that the $\psi_{n}^{L_0}(x)$ form a complete
set of {\it approximate} orthonormal eigenstates of $H$ if $L_0 \gg L$. The approximation is ever better, the larger is the ratio $L_0/L$.

\begin{figure}[b]
\centerline{
\includegraphics[width=\columnwidth]{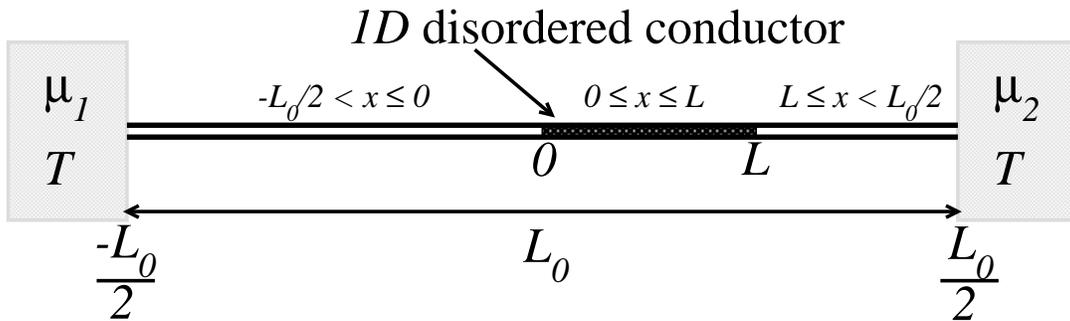}
}
\caption{
The system described by the wave function of Table \ref{structure of wf 2}: simulated is the effect of two reservoirs, placed at the ends of the interval $L_0$, which emit electrons and absorb, but do not reflect impinging electrons.
}
\label{1D and two reservoirs}
\end{figure}

\begin{figure}[h]
\centerline{
\includegraphics[width=\columnwidth]{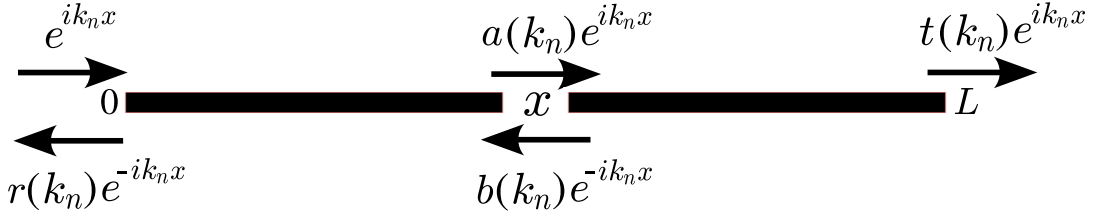}
}
\caption{
Schematic representation of the scattering problem for $\psi_{n>0}^{L_0}(x)$ of Table \ref{structure of wf 2}.
}
\label{Squematich_amplitudes_2}
\end{figure}

\begin{table}[h]
\tbl{The structure of the wave function $\psi_{n}^{L_0}(x)$ in the three regions shown in Fig. \ref{1D and two reservoirs}, when incidence is from the left, $n >0$ (2nd row), and when incidence is from the right, $n <0$ (3rd row). The wave function indicated inside the sample is that occurring between two successive scattering units. The various coefficients [$r(k_n)$, $a(k_n)$, etc.] are the same as those in Table \ref{structure of wf}, evaluated at $k=k_n$.}
{\begin{tabular}{| c | c | c | c | 
}
\hline
& $- L_0/2 < x \leq 0$
& $0 \leq x \leq L$ & $L \leq x < L_0/2$
\\
\hline\hline
$\psi_{n>0}^{L_0}(x)$
&
$\frac{1}{\sqrt{L_0}} \left(e^{ik_{n} x} + r(k_{n})e^{-ik_{n} x}\right)$
& $\frac{1}{\sqrt{L_0}} \left( a(k_n) e^{ik_{n} x} + b(k_n)  e^{-ik_{n} x} \right)$ & 
$\frac{1}{\sqrt{L_0}} \; t(k_{n}) e^{ik_{n} x}$
\\
\hline
$  \psi_{n<0}^{L_0}(x)      $
& $ \frac{1}{\sqrt{L_0}} \; t'(k_{n}) e^{-ik_{n} x}$
& $\frac{1}{\sqrt{L_0}} \left( a'(k_n) e^{ik_{n} x} + b'(k_n)  e^{-ik_{n} x} \right)$ & 
$\frac{1}{\sqrt{L_0}}\left(e^{-ik_{n} x} + r'(k_{n})e^{ik_{n} x}\right)$
\\
\hline\hline
\end{tabular}}
\label{structure of wf 2}
\end{table}
The wave functions $\psi_{n}^{L_0}(x)$ in the various regions have the structure shown in
Table \ref{structure of wf 2}
(see also Fig. \ref{1D and two reservoirs}). 
The quantities $a(k_n), b(k_n)$, $a'(k_n), b'(k_n)$ at $x$ denote the amplitudes inside the disordered region between two successive scattering units;
we are assuming that between individual scatterers there is a free potential region, with a `small' width, where the wave function can be written as it is indicated in the second column of Table \ref{structure of wf 2}: see Fig. \ref{Squematich_amplitudes_2}.

We should remark that, 
although the unperturbed wave functions $\phi_{n}\left( x\right)$, Eq. \eqref{phi s}, are periodic and orthonormal in $(-L_{0}/2,L_{0}/2)$, 
the perturbed wave functions $\psi_{n}^{L_0}(x)$ are not periodic;
also, {\it they are not orthonormal, since they are not
eigenstates with different eigenvalues of a Hermitean operator}.
However, as we mentioned, we found, in various cases, that the resulting wave functions $\psi_n^{L_0}(x)$ of Table \ref{structure of wf 2} fulfill
\begin{subequations}
\begin{eqnarray}
(\psi_n^{L_0}, \psi_{n'}^{L_0})_{L_0}
&\approx& (\phi_n, \phi_{n'})_{L_0}
= \delta_{n n'} , 
\label{(psi_s,psi_s')L' a}
\end{eqnarray}
where we have defined the scalar product
\begin{equation}
(\chi^{\prime \; L_0}, \chi^{L_0})_{L_0}
\equiv \int_{-L_0/2}^{L_0/2} \left[ \chi^{\prime \; L_0}(x)\right] ^{*}
\chi^{L_0}(x)dx;
\label{(,)L'}
\end{equation}
the approximate orthonormality is ever better the larger is the ratio $L_0/L$, i.e.,
\begin{eqnarray}
L_0/L \gg 1.
\label{L0} 
\end{eqnarray}
\label{(psi_s, psi_s')L'}
\end{subequations}

We have found wide analytical and numerical evidence to verify the approximation of Eq. \eqref{(psi_s, psi_s')L'}.
The result (\ref{(psi_s, psi_s')L'}) is not surprising: from the statement of Eqs. (\ref{(psi_s, psi_s')L'}) for {\it discrete} momenta $k_n$, giving orthonormality in terms of {\it Kronecker's delta}, one recovers, in the continuous limit, as $L_0 \to \infty$, orthonormality in the {\it Dirac $\delta$-function} sense. 

The situation we are considering, illustrated in Fig. \ref{1D and two reservoirs}, simulates the effect of two reservoirs placed at the ends of the interval $L_0$, which emit electrons, absorb but do not reflect back \cite{buettiker}, so they do not alter the 'cut-down' wave functions
of Eq. (\ref{cut-down wfs}). 
Such reservoirs are assumed to be at temperature $T$ and chemical potentials $\mu_1$ and $\mu_2$, respectively.

We thus see that, according to Eqs. (\ref{(psi_s, psi_s')L'}), the two perfect leads on both sides of the actual system $L$, introduced for 'technical reasons' by B\"uttiker \cite{buettiker}, must have the property $L_0/L \gg 1$.

\subsection{Explicit calculation of the electron density in the case of a disordered system}
\label{density_disordered_system}

We use the results of the previous subsection to extend the analysis of 
Sec. \ref{ballistic 1D}, which describes a ballistic conductor, to a disordered system.

We introduce creation and annihilation operators $d^{\dagger}_n$, $d_n$, which create or annihilate an electron in state $\psi^{L_0}_{k_{n}}(x)$ of Eq. (\ref{cut-down wfs}); i.e.,
\begin{eqnarray}
\psi^{L_0}_n(x) \Rightarrow d^{\dagger}_n | 0 \rangle
\equiv | 1_n  \rangle ,
\;\;\;\;\;
i.e., \;\;\;\;\;
\left\{
\begin{array}{cc}
\psi^{L_0}_{n>0}(x) & \;\;\; \Rightarrow \;\;\;  d^{\dagger}_{n>0} | 0 \rangle = | 1_{n>0}  \rangle ,  \\
\psi^{L_0}_{n=0} \equiv 0 , &  \\
\psi^{L_0}_{n<0}(x) & \Rightarrow d^{\dagger}_{n<0} | 0 \rangle = | 1_{n<0}  \rangle .
\end{array}
\right.
\label{sps perturbed}
\end{eqnarray}
The property of the states $\psi^{L_0}_n(x)$, Eq. (\ref{(psi_s,psi_s')L' a}), is inherited by the single-particle states written in second quantization, as
\begin{eqnarray}
(\psi^{L_0}_n(x), \psi^{L_0}_{n'}(x)) _{L_0} \approx \delta_{nn'}    \;\;\; 
\Rightarrow \;\;\; 
\langle 0 | d_n d^{\dagger}_{n'}| 0 \rangle = \delta_{n n'}.
\end{eqnarray}
As one further example, two-particle antisymmetric states are
\begin{eqnarray}
\frac{1}{\sqrt{2!}}  \left|
\begin{array}{cc}
\psi^{L_0}_{n}(x_1) & \psi^{L_0}_{n}(x_2) \\
\psi^{L_0}_{n'}(x_1) & \psi^{L_0}_{n'}(x_2)
\end{array}
\right|
 \;\;\; 
\Rightarrow \;\;\; 
| 1_{n} 1_{n'} \rangle = d^{\dagger}_{n}d^{\dagger}_{n'} |0\rangle ,
\label{2-part states perturbed}
\end{eqnarray}
etc.

The new reduced density matrix is obtained from Eqs. (\ref{model density matrix})
replacing $c_n$ by $d_n$.
The factor $n=0$ is omitted, as it corresponds to 
$\psi^{L_0}_{n=0}(x) \equiv 0$; i.e.,
\begin{subequations}
\begin{eqnarray}
\widehat{\rho}(\beta, \mu_1, \mu_2)
&=&  
\frac{  e^{-\beta \sum_{n > 0} (\epsilon_n - \mu_1) d^{\dagger}_n d_n  }}
{{\cal Z}^{(+)}(\beta, \mu_1)} 
\times 
\frac{  e^{-\beta \sum_{n < 0} (\epsilon_n - \mu_2) d^{\dagger}_n d_n  }}
{{\cal Z}^{(-)}(\beta, \mu_2)}  
\label{model density matrix disordered a} 
\\
&\equiv&  \widehat{\rho}_{+}(\beta, \mu_1) \;
\widehat{\rho}_{-}(\beta, \mu_2) ,  
\label{model density matrix disordered b} 
\end{eqnarray}
\begin{eqnarray}
{\cal Z}^{(+)}(\beta, \mu_1)
&=& \prod_{n > 0} \left[1 + e^{-\beta(\epsilon_n - \mu_1)}     \right] ,
\label{model density matrix disordered c} \\
{\cal Z}^{(-)}(\beta, \mu_2)
&=& \prod_{n < 0} \left[1 + e^{-\beta(\epsilon_n - \mu_2)}     \right]  .
\label{model density matrix disordered d}
\end{eqnarray}
\label{model density matrix disordered}
\end{subequations}

The above reduced density matrix for the system, 
Eqs. (\ref{model density matrix disordered}), is suggested here as a simple model
to describe a non-equilibrium, but stationary state.
Just as we mentioned in relation with the density matrix of 
Eq. (\ref{model density matrix}) for the ballistic case, 
in Ref.  \cite{dhar_saito_hanggi}
a non-equilibrium density-matrix
description of steady-state quantum transport
is presented. 
In App. \ref{dhar_et_al_density_matrix} we give the correspondence between the above ansatz, Eqs. (\ref{model density matrix disordered}), and the results of
Ref.  \cite{dhar_saito_hanggi}.

The current through a 1D disordered conductor and the associated conductance can be computed by using the above model for the reduced density matrix and one finds the well-known Landauer-B\"uttiker result \cite{buettiker}. Here we concentrate on the electron density, as it is the main purpose of this paper.

As before, by 'total system' we mean the actual system (sample) contained in the interval from $0$ to $L$, plus the leads;
thus the total system corresponds to the interval from $-L_0/2$ to $L_0/2$
(see Fig. \ref{1D and two reservoirs}).

The electron-density operator at the point $x$ for an $N$-electron system is given by
\begin{eqnarray}
\widehat{D}_{el}(x)
\equiv  \sum_{i=1}^{N} \delta(x-\hat{x}_i) . 
\label{density op dis}
\end{eqnarray}
Its matrix elements in the one-particle states $\psi^{L_0}_{n}(x)$ of Eqs. (\ref{(psi_s, psi_s')L'}) are
\begin{eqnarray}
{D}^{(1)}_{n n'}(x)
&\equiv& \left(\psi^{L_0}_{n}(x_1), \hat{D}_1(x) \psi^{L_0}_{n'}(x_1)\right)
= [\psi^{L_0}_{n}(x)]^{*} \; \psi^{L_0}_{n'}(x)    .       
\label{rho_ss' dis}
\end{eqnarray}

In a second quantization formalism, the electron-density operator at the point $x$ takes the form
\begin{eqnarray}
\widehat{\mathbb{D}}_{el}(x)
&=& \sum_{n n'} 
{D}^{(1)}_{nn'}(x) 
\; \hat{d}^\dagger_n \hat{d}_{n'}  \; .
\label{2nd quant density dis}
\end{eqnarray}
Its expectation value in the state defined by the density matrix $\hat{\rho}$
of Eqs. (\ref{model density matrix disordered}) is
\begin{subequations}
\begin{eqnarray}
{\cal W}(x)
&\equiv& {\rm Tr}[\hat{\rho} \; \hat{\mathbb{D}}_{el}(x)]  
\label{D=Tr rho W dis a}  \\
&=& \sum_{nn'}  D^{(1)}_{nn'}(x) \;
{\rm Tr} ( \hat{\rho} \; \hat{d}^\dagger_n \hat{d}_{n'} )   . 
\label{D=Tr rho W dis b}
\end{eqnarray}
\label{D=Tr rho W dis}
\end{subequations}
For the density matrix of Eq. \eqref{model density matrix disordered} one finds
\begin{eqnarray}
{\rm Tr} ( \hat{\rho} \; \hat{d}^\dagger_n \hat{d}_{n'})
=\delta_{nn'}{\rm Tr}(\hat{N}_n \hat{\rho}),
\label{tr rho d+d}
\end{eqnarray}
$\hat{N}_n = \hat{d}^\dagger_n \hat{d}_{n}$
being the number operator for state $n$.

In {\it equilibrium}, we have
\begin{subequations}
\begin{eqnarray}
{\rm Tr} (\hat{\rho}\hat{N}_n)
&=&\langle  \hat{N}_n \rangle_{\beta,\mu}    \\
&=&f_{\mu, \beta}(\epsilon_n)
= \frac{1}{1+ e^{\beta(\epsilon_n -\mu)}}  ,
\end{eqnarray}
\label{<rho Ns> a a}
\end{subequations}
$f_{\mu, \beta}(\epsilon_n)$ being the Fermi function.

For the {\it non-equilibrium stationary state}
defined by the model density matrix of Eqs. (\ref{model density matrix disordered}),
we have
\begin{subequations}
\begin{eqnarray}
{\rm Tr} (\hat{\rho}\hat{N}_{n^{+}})
&=&f_{\mu_1, \beta}(\epsilon_{n^{+}}) \\
{\rm Tr} (\hat{\rho}\hat{N}_{n^{-}})
&=&f_{\mu_2, \beta}(\epsilon_{n^{-}})
\end{eqnarray}
\label{<rho Ns> bb}
\end{subequations}
and Eqs. (\ref{rho_ss' dis}), (\ref{D=Tr rho W dis}), (\ref{tr rho d+d}) and (\ref{<rho Ns> bb})
give the general expression for the electron density in the interval 
$-L_0/2 < x < L_0/2$, i.e., in the leads and inside the system proper, as
\begin{subequations}
\begin{eqnarray}
{\cal W}(x)
 &=&\sum_{n}  D^{(1)}_{nn}  \;
{\rm Tr} ( \hat{\rho} \; \hat{N}_n )  
\label{W=Tr rho D 2 a}  \\
&=& \sum_{n > 0}  D^{(1)}_{nn}  \;
f_{\mu_1, \beta}(\epsilon_n)
+ \sum_{n<0} D^{(1)}_{nn}  \;
f_{\mu_2, \beta}(\epsilon_n)  
\label{W=Tr rho D 2 b}   \\
&=& \sum_{n>0} |\psi^{L_0}_{n}(x)|^2 f_{\mu_1, \beta}(\epsilon_n)
+\sum_{n<0} |\psi_{n}^{L_0}(x)|^2 f_{\mu_2, \beta}(\epsilon_n)
\label{W=Tr rho D 2 c} 
\end{eqnarray}
\label{W=Tr rho D 2}
\end{subequations}
In physical terms, this final result means that the density at $x$ is the sum of the densities 
of left-going and right-going electrons at $x$, both weighted by the probability of such an electron having been emitted from the appropriate reservoir.

In the next section, we shall find explicit expressions for the above result in the various regions. 
Subsequently, these results will be averaged over an ensemble of configurations of disorder and compared with computer simulations.

In the simulations, the disordered potential is a random function of position.
We use the model employed in Ref.  \cite{mello-yepez}, also used in Ref.  \cite{froufe_et_al_pre_2007}, which we now summarize.

In the model we shall use, the scattering units consist of thin potential slices idealized as equidistant delta potentials    
--$d$ being their separation--, 
labeled by the index $ j=1, \cdots, N_{scatt}$.
Thus, there are a total of $N_{scatt}$
such delta potentials in the sample of length $L$.
We shall always consider situations in which the wavelength involved is much larger than the separation between successive scatterers, i.e., $\lambda \gg d$.
The random potential, in units of $\hbar^2/2m$, has the form
\begin{subequations}
\begin{eqnarray}
u(x) &=& \sum_{j=1}^{N_{scatt}} u_j \delta(x-jd),
\\
{\rm with} \;\;\;\;\; L &=& N_{scatt} d. 
\end{eqnarray}
\label{MicrocopicPotentialModel}
\end{subequations}
The strength $u_j$ of a given delta potential is taken to be statistically independent from,
and identically distributed to, the strength of any other one;
therefore, the potential strengths $u_{j}$ (in units of $k$) are uniformly distributed in the interval $\left[-u_{0},u_{0}\right]$, $u_{0}$ being the maximum strength.

In the dense weak scattering limit, the average reflection coefficient for a single delta scatterer (indicated by the index $1$) 
\begin{equation}
\left\langle  R_1(k_n) \right\rangle = \langle |r_1(k_n)|^2\rangle = 
\left\langle 
\frac{\left( \frac{u_{1}}{2k_{n}}\right)^{2}}{1+\left( \frac{u_{1}}{2k_{n}}\right)^{2}}
\right\rangle ,
\label{Def_Coef_R_1}
\end{equation}
plays a key role.
In this limit, the potential strength of a 
delta scatterer is very weak and their linear density is very large, so that the resulting mean-free-path (mfp) is fixed (for each electron level, since it is level dependent, as explained below); the corresponding statistical properties of the full system depend only on the mfp and on no other property of the delta-potentials distribution \cite{froufe_et_al_pre_2007}.

Throughout the paper, the concept of mfp is defined as in Ref. \cite{froufe_et_al_pre_2007}.
In the present problem, it depends on the energy level $n$, so it is designated as $\ell_n$.
At the momentum $k_n$, when the average reflection coefficient
for a single scatterer 
$\langle R_1(k_n)\rangle \ll 1$, it is defined through the relation
\begin{eqnarray}
\frac{1}{\ell_n} &=& \frac{\langle R_1(k_n)\rangle}{d} .
\label{1/l}
\end{eqnarray}
The ratio $s_n \equiv L/\ell_n$, frequently employed in the paper, of the system length to the mfp is then given by 
\begin{equation}
s_n \equiv \frac{L}{\ell_n} 
= L\frac{\langle R_1(k_n)\rangle}{d}.
\label{L/l}
\end{equation}
The ratio $L/\ell_n$ decreases as we go up in the spectrum, since $\langle  R_1(k_n) \rangle$ decreases; the system becomes ever more ballistic and thus delocalized.

Definition (\ref{1/l}) is applicable as long as $\langle R_1(k_n)\rangle \ll 1$. As the incident energy is decreased, $\langle R_1(k_n)\rangle$ increases and the mfp decreases. Definition (\ref{1/l}) is not strictly applicable near the ground state, where $\langle R_1(k_n)\rangle \sim 1$; in that region, the quantity $s_n$, defined as 
\begin{equation}
s_n \equiv N_{scatt} \langle  R_1(k_n) \rangle,
\label{L/l_bis}
\end{equation}
will be taken as a useful parameter that measures the extent to which the wave function penetrates into the sample. The local-density-of-states (LDOS), being proportional to the intensity itself, is ever more depleted as we go down in energy.

The specific theoretical model that our computer simulations will be compared with is designated as the DMPK model \cite{mello-kumar,Dorokhov_1984,mpk}. This is essentially a random-phase approximation for a fixed energy $\epsilon_n$; the DMPK model is governed by a diffusion equation in the transfer-matrix space, which depends on the single parameter $s_n= L/\ell_n$ that was defined in Eq. (\ref{L/l}). 
For one dimensional systems, the DMPK equation reduces to Melnikov's \cite{Melnikov_1980}. 

Even though the DMPK model depends only on the microscopic details through the ratio $L/\ell_{n}$, the DMPK predictions are valid when the parameter $z_{n} \equiv k_{n}\ell_{n}$ satisfies the so called {\it weak disorder regime} \cite{Beenakker_1997}, i.e., 
\begin{equation}
z_{n} \equiv k_{n}\ell_{n}\gg1.
\label{kl_parameter}
\end{equation}
For the potential model used in our computer simulations, Eq. \eqref{MicrocopicPotentialModel}, and the definition of the mfp, Eq. \eqref{1/l}, the parameter $z_{n}$ is written as
\begin{equation}
z_{n} =\frac{k_{n}d}{\left\langle R_{1}\left(k_{n}\right)\right\rangle },\;\; \mathrm{with} \;\;k_{n}d\ll 1, \;\; \forall n.
\label{WeakDisorderCondition}
\end{equation}
The weak disorder condition, Eq. \eqref{kl_parameter}, is satisfied as long as $\langle R_1(k_n)\rangle \ll k_{n}d \ll 1$, i.e., for high-lying energy levels, with $n \gg 1$. As the incident energy decreases $\langle R_1(k_n)\rangle \approx 1$ and $k_{n}d \ll 1$, so the weak disorder condition is not fulfill for low-lying energy levels, with $n \sim 1$ near the ground state.

Equations \eqref{L/l_bis} and \eqref{WeakDisorderCondition} show that the average reflection coefficient for a single delta scatterer $\left\langle  R_1(k_n) \right\rangle $, Eq. \eqref{Def_Coef_R_1}, 
plays a key role to characterize our numerical simulations for both, high-lying and low-lying energy levels, and also to compare them with the DMPK predictions; therefore, the numerical results presented here will be specified by the parameters $s_{n}$, $z_{n}$ and $\left\langle  R_1(k_n) \right\rangle $: see Table \ref{Details_wn}.

\section{Explicit expressions for the electron density in the various regions
and their expectation value}
\label{analysis electron density}

In this section we obtain more explicit expressions for
the electron density in the various regions of the conductor, and compute their expectation value over an ensemble of configurations of disorder,
We shall restrict the analysis to the particular case of {\it zero temperature}.

\subsection{The density outside the sample}
\label{density_outside}

From Eq. (\ref{W=Tr rho D 2 c}) one finds, outside the sample, in the {\it left ballistic region},
$x\in [-L_0/2, 0]$, 
the general result
\begin{subequations}
\begin{eqnarray}
{\cal W}(x \in [-L_0/2,0])
&=&\frac{1}{L_0}\sum_{n>0}^{n_{max}^{+}}
\left| e^{ik_nx}+r(k_{n}) e^{-ik_nx} \right|^2 
+  \frac{1}{L_0}\sum_{n<0}^{n_{max}^{-}} \left| t'(k_{n}) e^{-ik_nx} \right|^2  \\
&\approx& \frac{1}{L_0}\sum_{n>0}^{\frac{N-\Delta N}{2}} 2
+ \frac{1}{L_0}\sum_{n>0}^{\frac{N+\Delta N}{2}}\Big(r(k_{n}) e^{-2i k_{n} x} + cc \Big)
+\frac{\Delta N}{L_0}
\left[
1+ R(\epsilon_F)
\right] \; .
\nonumber \\
\end{eqnarray}
\label{W(x in -L'/2 , 0) 1}
\end{subequations}
Here, $n_{max}^{+} > 0$ denotes the quantum number $n$ associated with the highest level fed by the left reservoir with right-going electrons, and
$n_{max}^{-} < 0$ is the quantum number $n$ associated with the highest level fed by the right reservoir with left-going electrons.
In the present model, at $T=0$ 
the number of electrons traveling to the left and the number of electrons traveling to the right are fixed.
We have defined
\begin{subequations}
\begin{eqnarray}
 && n^{+}_{max} +|n^{-}_{max}| =N    ,    \\
&&  n^{+}_{max} - |n^{-}_{max}| = \Delta N ,
\end{eqnarray}
\end{subequations}
$N$ being the total number of electrons, so that
\begin{subequations}
\begin{eqnarray}
 && n^{+}_{max} = \frac{N+\Delta N}{2} ,       \\
&&  |n^{-}_{max}| = \frac{N-\Delta N}{2}.
\end{eqnarray}
\end{subequations}
We also recall that we have defined $k_n$ as a positive number throughout the whole analysis.

In the {\it right ballistic region},
$x \in [L, L_0/2]$,
\begin{subequations}
\begin{eqnarray}
{\cal W}(x \in [L, L_0/2])
&=&\frac{1}{L_0}\sum_{n<0}^{n_{max}^{-}}
\left| e^{-ik_{n}x}+r'(k_{n}) e^{ik_{n}x} \right|^2 
+  \frac{1}{L_0}\sum_{n>0}^{n_{max}^{+}} \left| t(k_{n}) e^{ik_{n}x} \right|^2  \\
&\approx& \frac{1}{L_0}\sum_{n>0}^{\frac{N-\Delta N}{2}} 
\Big[2 + \Big(r'(k_{n}) e^{2i k_{n} x} + cc\Big)\Big]  
+\frac{\Delta N}{L_0} T(\epsilon_F) .
\end{eqnarray}
\label{W(x in L, L'/2) 2}
\end{subequations}
Here and in the previous equations, $R(\epsilon_F) = |r(\epsilon_F)|^2$ 
and $T(\epsilon_F) = |t(\epsilon_F)|^2$ are the reflection and transmission coefficients,  respectively.

The average over an ensemble of disorder configurations gives the general results
\begin{subequations}
\begin{eqnarray}
\Big\langle {\cal W}(x \in [-L_0/2,0])\Big\rangle
&=&  {\cal W}_0 
+\frac{\Delta N}{L_0}
\left\langle  R(\epsilon_F) \right\rangle 
+ \frac{1}{L_0}\sum_{n>0}^{\frac{N+\Delta N}{2}}\Big(\langle r(\epsilon_{n})\rangle e^{-2i k_{n} x} + cc \Big)
\nonumber \\
\label{<W(x in -L'/2 , 0)> 2}
\end{eqnarray}
and
\begin{eqnarray}
\Big\langle {\cal W}(x \in [L,  L_0/2])\Big\rangle
&=& {\cal W}_0
-\frac{\Delta N}{L_0}\langle R(\epsilon_F) \rangle 
+ \frac{1}{L_0}\sum_{n>0}^{\frac{N-\Delta N}{2}} 
\Big[\langle r'(k_{n})\rangle e^{2i k_{n} x} + cc \Big]  
\label{<W(x in 0, L'/2 >  b 2} \; ,
\nonumber \\
\end{eqnarray}
\label{<W(x in 0, L'/2 > 2}
\end{subequations}
where we have written ${\cal W}_0 = N/L_0$ for the electron density in the absence of disorder.

{\it In the DMPK approximation} \cite{mello-kumar,mpk}, the last terms in Eqs.
(\ref{<W(x in -L'/2 , 0)> 2})  and  (\ref{<W(x in 0, L'/2 >  b 2}) vanish and
\begin{subequations}
\begin{eqnarray}
\Big\langle
{\cal W}(x \in [-L_0/2,0])
\Big\rangle_{DMPK}
&=&{\cal W}_0 
+\frac{\Delta N}{L_0}\left\langle R(\epsilon_F) \right\rangle
\label{<W(x)> outside left in DMPK 1}\\
\Big\langle
{\cal W}(x \in [L,L_0/2])
\Big\rangle_{DMPK}
&=&{\cal W}_0 
-\frac{\Delta N}{L_0}\left\langle R(\epsilon_F) \right\rangle
\label{<W(x)> outside right in DMPK 1} .
\end{eqnarray}
\label{<W(x)> outside in DMPK 1}
\end{subequations}

\subsection{The density inside the sample: $x\in[0,L]$}
\label{density_inside}

Inside the system proper, Eq. (\ref{W=Tr rho D 2 c}) gives the general result (see App. \ref{proof eq Wn+})
\begin{eqnarray}
{\cal W}(0 \leq x \leq L)
&=& \frac{1}{L_0}
\left\{
\sum_{n>0}
\Big| 
\alpha_2^{*}(\epsilon_n){\rm e}^{ik_{n}x} - \beta_2^{*}(\epsilon_n){\rm e}^{-ik_{n}x}
\Big|^2 T(\epsilon_n) f_{\mu_1 \beta}(\epsilon_n)
\right.
\nonumber \\
&& \;\;\;\; \left. 
+ \sum_{n<0}
\Big| 
\beta_1(\epsilon_n){\rm e}^{ik_{n}x} +\alpha_1^{*}(\epsilon_n){\rm e}^{-ik_{n}x}
\Big|^2 T(\epsilon_n) f_{\mu_2 \beta}(\epsilon_n)
\right\},
\label{W(x) inside}
\end{eqnarray}
which contains the contribution of the electrons that impinge on the system from the left with positive momentum $k_{n}$, and from the right with negative momentum $-k_{n}$. Here, $T(\epsilon_n) = |t(\epsilon_n)|^2$ is the transmission coefficient. Just as above, we consider the {\it zero-temperature} limit, $T=0$, while the two chemical potentials will be taken to be different, the left one being higher, $\mu_1>\mu_2$.

The expectation value of the  electron density of Eq. (\ref{W(x) inside}) over an ensemble of configurations of disorder is given by
\begin{subequations}
\begin{eqnarray}
L_0\left\langle
{\cal W}(x \in [0,L])
\right\rangle
&=&
\sum_{n>0}^{n^{+}_{max}}
\left\langle
\Bigg|
\alpha_2^{*}(\epsilon_n){\rm e}^{ik_{n}x} - \beta_2^{*}(\epsilon_n){\rm e}^{-ik_{n}x}
\Big|^2 T(\epsilon_n)
\right\rangle 
\nonumber \\
&&\hspace{5cm}+\sum_{n<0}^{n^{-}_{max}}
\left\langle\Big| 
\beta_1(\epsilon_n){\rm e}^{ik_{n}x} +\alpha_1^{*}(\epsilon_n){\rm e}^{-ik_{n}x}
\Big|^2 T(\epsilon_n)
\right\rangle    
\nonumber \\ \\
&=& \sum_{n=0}^{|n^{-}_{max}|}
\Bigg[
\left\langle
\Big|
\alpha_2^{*}(\epsilon_n){\rm e}^{ik_{n}x} - \beta_2^{*}(\epsilon_n){\rm e}^{-ik_{n}x}
\Big|^2 T(\epsilon_n)
\right\rangle
\nonumber \\ 
&&\hspace{5cm}+\left\langle\Big| 
\beta_1(\epsilon_n){\rm e}^{ik_{n}x} +\alpha_1^{*}(\epsilon_n){\rm e}^{-ik_{n}x}
\Big|^2 T(\epsilon_n)
\right\rangle 
\Bigg]
\nonumber \\
&&+\sum_{ n=|n^{-}_{max}| }^{ n^{+}_{max} }
\left\langle
\Bigg|
\alpha_2^{*}(\epsilon_n){\rm e}^{ik_{n}x} - \beta_2^{*}(\epsilon_n){\rm e}^{-ik_{n}x}
\Big|^2 T(\epsilon_n)
\right\rangle  
\label{dens inside general 1 b}
\end{eqnarray}
\label{dens inside general 1}
\end{subequations}

1) E.g., in {\it equilibrium}, $\mu_1=\mu_2$,
\begin{eqnarray}
L_0\left\langle
{\cal W}(x \in [0,L])
\right\rangle
&=& \sum_{n=0}^{n^{+}_{max}=|n^{-}_{max}|=\frac{N}{2}}
\Bigg[
\left\langle
\Big|
\alpha_2^{*}(\epsilon_n){\rm e}^{ik_{n}x} - \beta_2^{*}(\epsilon_n){\rm e}^{-ik_{n}x}
\Big|^2 T(\epsilon_n)
\right\rangle
\nonumber           \\
&&\hspace{5cm}  +\left\langle\Big| 
\beta_1(\epsilon_n){\rm e}^{ik_{n}x} +\alpha_1^{*}(\epsilon_n){\rm e}^{-ik_{n}x}
\Big|^2 T(\epsilon_n)
\right\rangle
\Bigg]   
\nonumber \\
\label{general, equil}
\end{eqnarray}

We compute the above expectation value of the  electron density over an ensemble of configurations of disorder in the {\it DMPK approximation} \cite{mello-kumar,mpk}, following the procedure of Ref.  \cite{{mello-shi-genack}}. We find
\begin{subequations}
\begin{eqnarray}
&&\left\langle 
\Big|
\alpha_2^{*}(\epsilon_n){\rm e}^{ik_{n}x} 
-\beta_2^{*}(\epsilon_n){\rm e}^{-ik_{n}x} 
\Big|^2 T(\epsilon_n)
\right\rangle_{DMPK} =
1- \int_0^{\infty}\int_0^{\infty} g(\lambda_1, \lambda_2)
p_{s_{n}^{(1)}}(\lambda_1)p_{s_{n}^{(2)}}(\lambda_2) d\lambda_1 d\lambda_2 
\nonumber \\
\label{direct calc left incidence}       \\
&&\left\langle 
\Big|
\beta_1(\epsilon_n){\rm e}^{ik_{n}x} + 
\alpha_1^{*}(\epsilon_n){\rm e}^{-ik_{n}x} 
\Big|^2 T(\epsilon_n) \right\rangle_{DMPK} 
= 1+ \int_0^{\infty}\int_0^{\infty} g(\lambda_1, \lambda_2)
p_{s_{n}^{(1)}}(\lambda_1)p_{s_{n}^{(2)}}(\lambda_2) d\lambda_1 d\lambda_2 \; ,
\nonumber \\
\label{direct calc right incidence}      
\end{eqnarray}
\label{W(0- n DMPK)}
\end{subequations}
where
\begin{subequations}
\begin{eqnarray}
g(\lambda_1, \lambda_2)
=\frac{\lambda_1 - \lambda_2}{1+\lambda_1 + \lambda_2} \; ,
\\
s_{n}^{(1)}\equiv\frac{x}{\ell_{n}}, \;\;\;\;\;\; s_{n}^{(2)}\equiv\frac{L-x}{\ell_{n}}.
\end{eqnarray}
\label{}
\end{subequations}
Here, the index $i=1,2$ refers to the fraction of the wire on the left and on the right of the observation point $x$, respectively. The parameter $\lambda_{i} \ge 0$ is one of the variables (in addition to two phases) defining a transfer matrix for one open channel, and $p_{s^{(i)}}(\lambda_{i})$ is its statistical distribution given by DMPK for a specific value of $s_{n}^{(i)}$ (for one open channel, the DMPK equation reduces to Melnikov's \cite{Melnikov_1980}).
The transmission coefficient is given in terms of $\lambda$ as $T=1/(1+\lambda)$.

To illustrate the meaning of the probability density $p_s(\lambda)$, we give in App. \ref{1st and 2nd mom of lambda} Melnikov's equation and the first and second moments of $\lambda$ associated with such a distribution, where $s=L/\ell$, $\ell$ being the transport elastic mean free path, which is the only microscopic parameter in the DMPK formalism.

In the sum (\ref{general, equil}), each term with
$n>0$ has its counterpart for $n<0$;
for one given $\epsilon_n$ we then have, due to Eq. \eqref{W(0- n DMPK)}
\begin{eqnarray}
\left\langle
\Big|
\alpha_2^{*}(\epsilon_n){\rm e}^{ik_{n}x} - \beta_2^{*}(\epsilon_n){\rm e}^{-ik_{n}x}
\Big|^2 T(\epsilon_n)
\right\rangle_{DMPK}
+\left\langle\Big| 
\beta_1(\epsilon_n){\rm e}^{ik_{n}x} +\alpha_1^{*}(\epsilon_n){\rm e}^{-ik_{n}x}
\Big|^2 T(\epsilon_n)
\right\rangle_{DMPK} =2
\label{2 for DMPK 1}
\nonumber \\
\end{eqnarray}
independent of $x$, and so ($N$= total number of electrons inside $L_0$)
\begin{eqnarray}
\left\langle
{\cal W}(x \in [0,L])
\right\rangle_{DMPK} 
&=&\frac{1}{L_0}\frac{N}{2} 2 = \frac{N}{L_0} 
\equiv {\cal W}_0 \; ,
\end{eqnarray}
within the DMPK model.
Notice that we do not obtain the result reported in Ref.  \cite{mello-shi-genack}, because the contributions of electrons traveling in both directions compensate to give a constant value.

   2) We go back to the case where {\it the two chemical potentials are not equal}, the left one being higher, $\mu_1>\mu_2$.
We do not have an equilibrium state, but a stationary state:
from Eq. (\ref{dens inside general 1 b})
\begin{eqnarray}
&&L_0\left\langle
{\cal W}(x \in [0,L])
\right\rangle_{DMPK}= 2 |n^{-}_{max}|
\nonumber
\\
&&+\sum_{ n=|n^{-}_{max}| }^{ n^{+}_{max} }
\left\langle
\Bigg|
\alpha_2^{*}(\epsilon_n){\rm e}^{ik_{n}x} - \beta_2^{*}(\epsilon_n){\rm e}^{-ik_{n}x}
\Big|^2 T(\epsilon_n)
\right\rangle
\nonumber \\
\label{W DMPK inside 1 d}
\end{eqnarray}
We have used the fact that there are $|n^{-}_{max}|$ terms in the first part of Eq. 
\eqref{dens inside general 1 b}, each giving a contribution of 2, which in turn arises from the DMPK approximation, as shown in Eq. (\ref{2 for DMPK 1}).
Recalling that $N-\Delta N = 2|n_{max}^{-}|$, we then have
\begin{subequations}
\begin{eqnarray}
L_0\left\langle
{\cal W}(x \in [0,L])
\right\rangle_{DMPK}
&\approx &(N-\Delta N) 
+ \Delta N 
\left\langle
\left|
\alpha_2^{*}(\epsilon_F){\rm e}^{ik_{F}x} - \beta_2^{*}(\epsilon_F){\rm e}^{-ik_{F}x}
\right|^2 T(\epsilon_F)
\right\rangle 
\label{<W> inside, DMPK 1 a} 
\nonumber   \\ \\
& = & N + \Delta N 
\left[\left\langle
\left|
\alpha_2^{*}(\epsilon_F){\rm e}^{ik_{F}x} - \beta_2^{*}(\epsilon_F){\rm e}^{-ik_{F}x}
\right|^2 T(\epsilon_F)
\right\rangle -1 \right] 
\label{<W> inside, DMPK 1 b}
\nonumber \\ \\
&=& N-\Delta N \int_0^{\infty}\int_0^{\infty} g(\lambda_1, \lambda_2)
p_{s_{F}^{(1)}}(\lambda_1)p_{s_{F}^{(2)}}(\lambda_2) d\lambda_1 d\lambda_2 
\nonumber \\
\label{<W> inside, DMPK 1 c}  \\
\left\langle
{\cal W}(x \in [0,L])
\right\rangle_{DMPK}
&=&{\cal W}_0
-\frac{\Delta N}{L_0} \int_0^{\infty}\int_0^{\infty} g(\lambda_1, \lambda_2)
p_{s_{F}^{(1)}}(\lambda_1)p_{s_{F}^{(2)}}(\lambda_2) d\lambda_1 d\lambda_2 .
\label{<W> inside, DMPK 1 d}
\nonumber \\
\end{eqnarray}
\label{<W> inside, DMPK 1}
\end{subequations}
\noindent Here, $s_{F}^{(1)}=x/\ell_{F}$ and $s_{F}^{(2)}=(L-x)/\ell_{F}$, $\ell_{F}$ being the mean free path at the Fermi level.

\section{The equilibrium density at zero temperature and its expectation value over disorder}
\label{expect_density_dmpk_simul}

In the present section we consider the disordered system in {\it equilibrium} at $T=0$, i.e., with no chemical potential difference ($\mu_{1}=\mu_{2}= \mu$) between the two reservoirs.
At $T=0$, the total number of electrons is fixed 
and equal to $N$.
The Fermi levels of the electrons traveling to the left and to the right, 
and the corresponding number of electrons, $n_{max}^{+}$ and $| n_{max}^{-} |$,
are equal, i.e.,
\begin{equation}
k_{F}=\frac{2\pi n_{max}^{+}}{L_{0}}=\frac{2\pi | n_{max}^{-} |}{L_{0}},
\;\;\;
n_{max}^{+} = |n_{max}^{-}| = \frac{N}{2}.
\end{equation}
In equilibrium, the electron density can be written as
\begin{equation}
\frac{{\cal W}(x)}{{\cal W}_0}
= \frac{2}{N} \sum_{n=1}^{N/2}  w_{n}(x),
\label{a}
\end{equation}
where ${\cal W}_{0} = N/L_{0} $ is the ballistic electronic density, and
\begin{equation}
w_n(x) \equiv \frac{1}{2}\left[ w_n^{LI}(x) + w_n^{RI}(x) \right],
\label{Individual_level_w_n}
\end{equation}
represents the {\it dimensionless density} for the $n$-th level (here and in what follows, $n$ is understood to be a {\it positive} number), which is the sum of the $n$-th level contributions for left incidence,
\begin{subequations}
\begin{equation}
w_{n}^{LI}(x) \equiv L_{0}|\psi^{L_{0}}_{n}\left( x \right)|^{2},
\label{wn LI}	
\end{equation}
and right incidence, 
\begin{equation}
w_{n}^{RI}(x) \equiv L_{0}|\psi^{L_{0}}_{-n}\left( x \right)|^{2}.\label{wn RI}
\end{equation}
\label{wn_LI_RI}
\end{subequations}

From table \ref{structure of wf 2}, in the ballistic regions
the dimensionless density of the $n$-th level, Eq. \eqref{Individual_level_w_n}, is written as
\begin{subequations}
\begin{eqnarray}
w_{n}(x) = 1+ Re\left[ r\left( k_{n}\right) e^{-2ik_nx}\right]  , 
&\;\;& -\frac{L_{0}}{2} < x <0 ,
\label{wn x<0}
\\
w_{n}(x) = 1+ Re\left[ r^{\prime}\left( k_{n}\right) e^{2ik_nx}\right] ,  
&\;\;& L < x < \frac{L_{0}}{2},
\label{wn x>L}
\end{eqnarray}
while inside the disordered system, Eq. \eqref{W(x) inside}, is given by
\begin{equation}
w_{n}(x) 
=\frac{1}{2}
\Biggl[
\Big| 
\alpha_2^{*}(\epsilon_n){\rm e}^{ik_{n}x} - \beta_2^{*}(\epsilon_n){\rm e}^{-ik_{n}x}
\Big|^2 T(\epsilon_n)
+
\Big| 
\beta_1(\epsilon_n){\rm e}^{ik_{n}x} +\alpha_1^{*}(\epsilon_n){\rm e}^{-ik_{n}x}
\Big|^2 T(\epsilon_n) 
\Biggr], 
\;\; 0 < x <  L .
\label{mn(x)inside}
\end{equation}
\label{All_wn x}
\end{subequations}
From Eqs. \eqref{a}-\eqref{All_wn x}, the expectation value over an ensemble of disorder configurations of the equilibrium electron density,
$\left\langle {\cal W}(x) /{\cal W}_{0} \right\rangle $, 
can be found from the average of the density 
$\left\langle w_{n}(x) \right\rangle $ for the individual $N/2$ levels.

The theoretical model that we shall compare with computer simulations is that provided by DMPK{ \cite{mello-kumar,mpk}.

\subsection{Individual levels}
\label{density wn}

{\it Outside} the system, $-L_{0} / 2 < x < 0$ and $L<x<L_{0}/2$, the DMPK theoretical prediction for $\left\langle w_{n}(x) \right\rangle $ is found by averaging 
Eqs. \eqref{wn x<0} and \eqref{wn x>L}.
Due to the random phase approximation considered in the DMPK model, this approach predicts a null value for the average of the reflection amplitudes of the disordered system, i.e., 
\begin{equation}
\left\langle  r\left( k_{n}\right) \right\rangle_{DMPK} = \left\langle  r ^{\prime}\left( k_{n}\right) \right\rangle_{DMPK}=0, \;\;\; \forall n;
\label{RandmPhasesApprox}
\end{equation}
therefore, outside the system $\left\langle  w_n(x) \right\rangle_{DMPK} = 1$.

{\it Inside} the system, $0<x<L$, the DMPK theoretical prediction for 
$\left\langle w_{n}(x) \right\rangle $ is found from Eqs. \eqref{2 for DMPK 1} and \eqref{mn(x)inside}. The result is also $\left\langle  w_n(x) \right\rangle_{DMPK} =1$, 
i.e., insensitive to disorder: this is a property arising from the DMPK model.

In summary, the DMPK prediction for the average density for any individual 
level is written as
\begin{equation}
\left\langle  w_n(x) \right\rangle_{DMPK} 
=\frac{1}{2}\left[
\left\langle w_{n}^{LI}(x)\right\rangle_{DMPK}
+\left\langle w_{n}^{RI}(x)\right\rangle_{DMPK}
\right]
=1,\; \forall n \;  \mathrm{and} \; \; \forall x.
\label{total density level n}
\end{equation}

\subsubsection{Numerical results and the DMPK prediction}
\label{density_wn_Sim_DMPK}

In order to analyze the DMPK prediction given in Eq. \eqref{total density level n}, we carry out, for individual levels, numerical simulations to obtain the ensemble average of dimensionless density $\left\langle  w_n(x) \right\rangle$. The simulations are done from high-lying energy levels, $n\gg 1$ to low-lying energy levels, $n \sim 1$; the maximum energy level considered for all simulations is $n_{max}^{+}=\vert n_{max}^{-} \vert=N/2=10000$, which corresponds to the Fermi level. For a given energy level $n$, the simulation considers an ensemble of $10^6$ realizations of disordered samples generated with the random potential model of Eq. \eqref{MicrocopicPotentialModel}; each disorder configuration consists of $N_{scatt}=8000$ random delta scatterers.

The numerical results found for $\left\langle w_{n}(x) \right\rangle$ are characterized by the parameters $s_{n}=L/\ell_{n}$ and $z_{n}=k_{n}\ell_{n}$, which in turn depend on the average reflection coefficient of a single delta scatterer $\langle R_1(k_n) \rangle$: see Eqs. \eqref{L/l} - \eqref{Def_Coef_R_1}. 
The simulations are also analyzed by using the average reflection coefficient of the sample $\left\langle R\left(k_{n}\right) \right\rangle$ and the average reflection amplitude of the sample $\left\langle r\left(k_{n}\right) \right\rangle$. These quantities allow us to know if the numerical results are in the localized regime ($L/\ell_{n}\gg 1$ and $\left\langle R\left(k_{n}\right) \right\rangle \simeq 1$), or if those satisfy the weak disorder condition ($k_{n}\ell_{n}\gg 1$) and the random phase approximation ($\left\langle r\left(k_{n}\right) \right\rangle \simeq 0$) of the DMPK model; therefore, in table \ref{Details_wn} we present the relevant details to analyze the numerical simulations for $\left\langle w_{n}(x) \right\rangle$.

\begin{table}[t]
\tbl{Relevant details of the numerical simulations shown in Sec. \ref{density_wn_Sim_DMPK} for the average dimensionless density $\left\langle  w_n(x) \right\rangle$. The parameters $\left\langle R_{1}\left(k_{n}\right) \right\rangle$, $s_{n}=L/\ell_{n}$, $z_{n}=k_{n}\ell_{n}$, $\left\langle R\left(k_{n}\right) \right\rangle$ and $\left\langle r\left(k_{n}\right) \right\rangle$ are presented for those energy levels $n$, whose simulations are discussed in Sec. \ref{density_wn_Sim_DMPK}.
}
{\begin{tabular}{| c | c | c | c | c | c |
}
\hline
$n$ & $\left\langle R_{1}\left(k_{n}\right) \right\rangle$ & $s_{n}=L/\ell_{n}$ & $z_{n}=k_{n}\ell_{n}$ & $\left\langle R\left(k_{n}\right) \right\rangle$ & $\left\langle r\left(k_{n}\right) \right\rangle$
\\
\hline\hline
$10000$ & $1.02\times 10^{-3}$ & $8.13$ & $61.6$ & $0.9795$ & $2.6 \times 10 ^{-4} - i 7.9 \times 10 ^{-3}$
\\
\hline
$9000$ & $1.23\times 10^{-3}$ & $8.22$ & $55.04$ & $0.9896$ & $1.4 \times 10 ^{-3} - i 1.1 \times 10 ^{-3}$
\\
\hline
$7000$ & $2.07\times 10^{-3}$ & $16.58$ & $21.7$ & $0.9987$ & $-1.8 \times 10 ^{-3} - i 2.4 \times 10 ^{-2}$
\\
\hline
$5500$ & $3.35\times 10^{-3}$ & $26.80$ & $10.31$ & $0.9999$ & $-8.9 \times 10 ^{-3} - i 4.7 \times 10 ^{-2}$
\\
\hline
$5000$ & $4.05\times 10^{-3}$ & $32.38$ & $7.76$ & $0.9999$ & $-1.3 \times 10 ^{-2} - i 5.9 \times 10 ^{-2}$
\\
\hline
$1500$ & $2.07\times 10^{-2}$ & $334.92$ & $0.2249$ & $1.0000$ & $-0.40  - i 0.21$
\\
\hline
$10$ &  $0.962$ & $7774.8$ & $ 6 \times 10^{-5} $ & $ 1.0000$ & $-0.995 - i 3 \times 10 ^{-3}$
\\
\hline\hline
\end{tabular}}
\label{Details_wn}
\end{table}

In Fig. \ref{density for L and for R incid c 1}, computer simulations are compared with Eq. \eqref{total density level n}. 
The comparison is done for a high-lying energy level, $n=9000 \gg 1$. In this case, the system is localized, the weak disorder condition is satisfied and the random phase assumption is a reasonable approximation: see table \ref{Details_wn}.
The main figure shows excellent agreement between simulations and the DMPK prediction inside the disordered region. The insets represent in more detail the results outside the sample, where the simulations show small oscillations around the DMPK result, Eq. \eqref{total density level n}; this effect is due to the small, but non-zero value of $\left\langle r\left(k_{n}\right) \right\rangle $, for left incidence, and $\left\langle r^{\prime}\left(k_{n}\right) \right\rangle $, for right incidence, when the energy level $n\gg 1$ is close to the Fermi level \cite{yepez-saenz}.

\begin{figure}[h]
\includegraphics[width=\columnwidth]{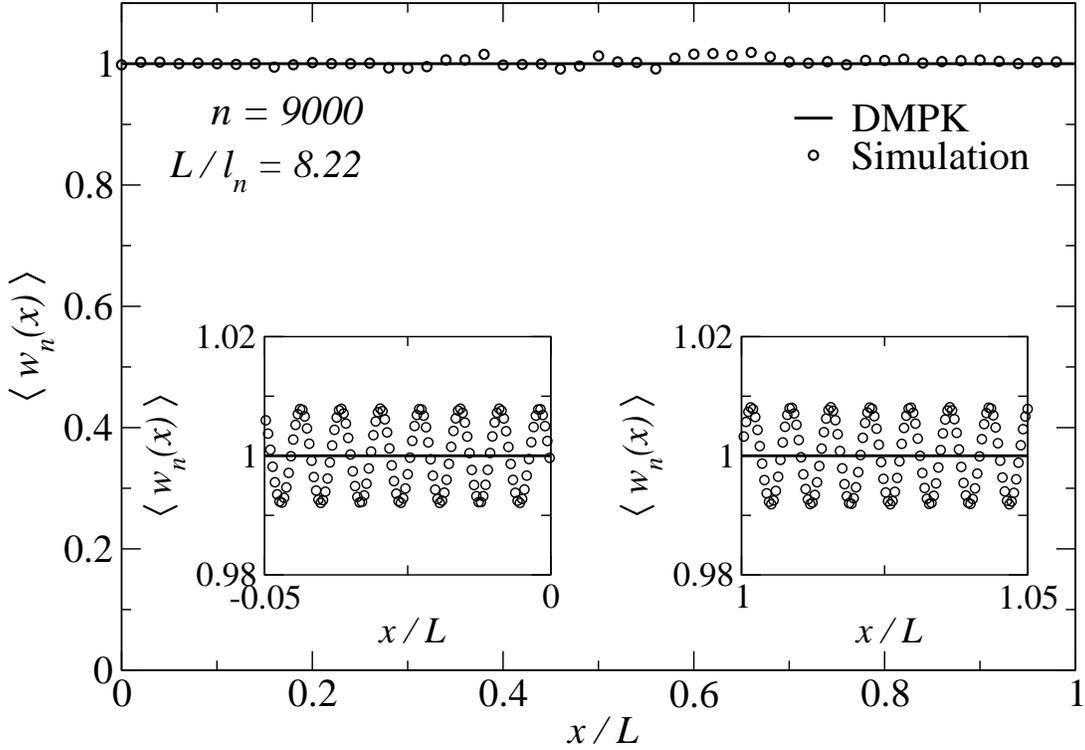}
\caption{The quantity $\left\langle w_n(x) \right\rangle$
from computer simulations, compared with 
the DMPK prediction, Eq. (\ref{total density level n}).
The system is in equilibrium at $T=0$, with the same chemical potential at both reservoirs.
The insets represent in more detail the results outside the sample.
The sample consists of $10^6$ realizations.
This figure repeats Fig. \ref{density for L and for R incid aa}a
below in more detail.
}
\label{density for L and for R incid c 1}
\end{figure}

Figure \ref{density for L and for R incid aa} shows, for six different energy levels, the result of computer simulations for the expectation value of the density for left incidence $\left\langle w_{n}^{LI}(x)\right\rangle$, right incidence $\left\langle w_{n}^{RI}(x)\right\rangle$, and their sum: $2\left\langle  w_n(x) \right\rangle=\left\langle w_{n}^{LI}(x)\right\rangle + \left\langle w_{n}^{RI}(x)\right\rangle$.

Panel \ref{density for L and for R incid aa}a) shows, for the maximum energy level $n_{max}^{+}=10000$, the numerical result for $2\left\langle w_{n}(x) \right\rangle$. For this high-lying energy level, the system is localized, the weak disorder condition is satisfied, and the random phase assumption of the DMPK model, Eq. \eqref{RandmPhasesApprox}, is a suitable approximation: see table \ref{Details_wn}. The comparison between the numerical result for $2\left\langle w_{n}(x) \right\rangle$ and the DMPK prediction given in Eq. \eqref{total density level n} shows an excellent agreement.

Panel \ref{density for L and for R incid aa}b) shows, for the high-lying energy level $n=7000$, the numerical results for $2\left\langle  w_{n}(x) \right\rangle$. The numerical results are in good agreement with the DMPK prediction given in Eq. \eqref{total density level n}. From table \ref{Details_wn}, the system is localized, the weak disorder requirement is satisfied, and 
the random phase assumption of the DMPK model is a suitable approximation for this high-lying energy level: see Eq. \eqref{RandmPhasesApprox}.
In this panel, three observations are in order: first, individual contributions to the electron density for left incidence $\left\langle  w_{n}^{LI}(x) \right\rangle$, and for right incidence $\left\langle  w_{n}^{RI}(x) \right\rangle$, banish before reaching the other end, so the electrons are penetrating less into the sample; second, the density appears to be constant, just as in panel \ref{density for L and for R incid aa}a); finally, although small, fluctuations appear in the center of the sample signaling that the weak disorder condition starts to get lost.

\begin{figure}[t]
\includegraphics[width=\columnwidth]{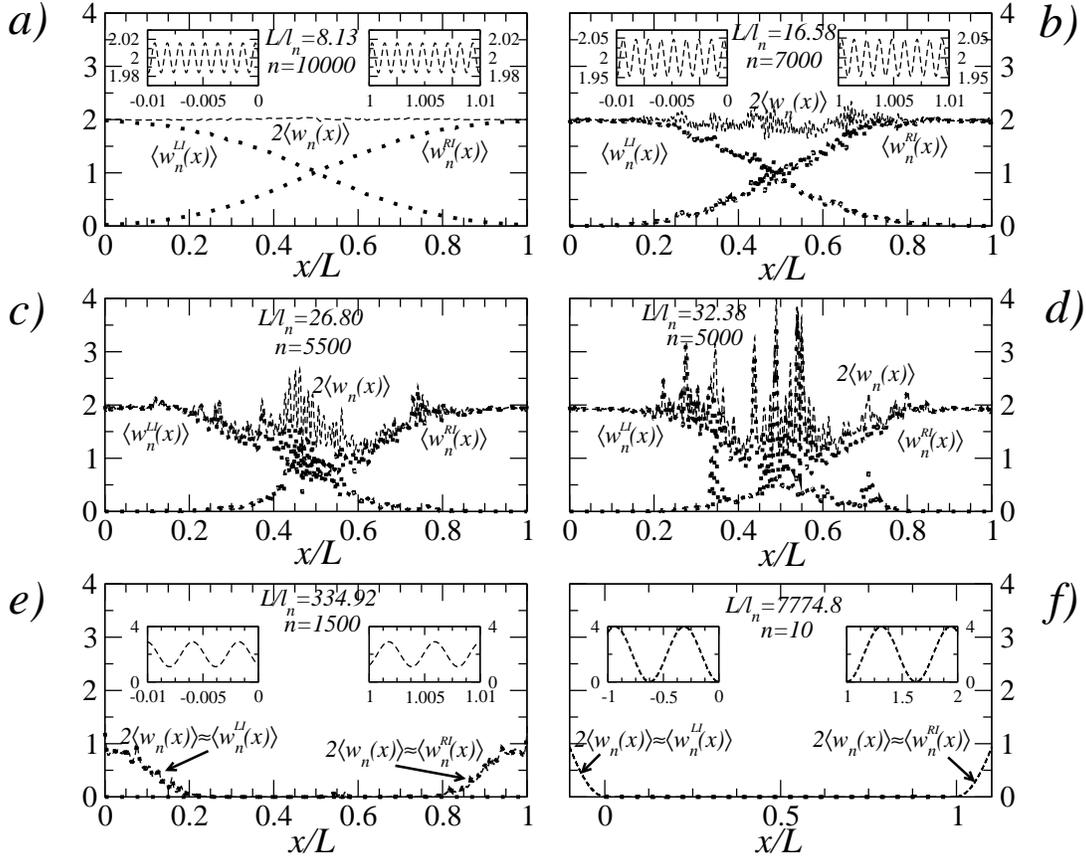}
\caption{Results of computer simulations for the electron density inside and outside a disordered 1D conductor for left incidence, right incidence and their sum ($\equiv 2w_n(x)$), for six individual energy levels identified by their $n>0$ value.
The value of the level dependent $s_n=L/\ell_n$ is indicated in each panel. 
The sum for both incidences is $x$-independent inside the sample for a rather wide range of $n$s.
For smaller $n$s, that sum, inside the sample, tends to be more and more concentrated near the edges $x=0,L$.
The insets in panels a), b), e), and f) show in detail the behavior of the electron density 
in a small region outside the sample.
}
\label{density for L and for R incid aa}
\end{figure}

Panels \ref{density for L and for R incid aa}c) and \ref{density for L and for R incid aa}d) show, respectively, the profiles of $2\left\langle  w_{n}(x) \right\rangle$ for the energy levels $n=5500$ and $n= 5000$. 
For these '{\it intermediate}' energy levels, the individual contributions to the electron density $\left\langle w_{n}^{LI}(x) \right\rangle$ for left incidence, and $\left\langle  w_{n}^{RI}(x) \right\rangle$ for right incidence, banish near the center of system; therefore, for both incidences, the electrons are penetrating less into the sample, giving rise to a drop at the center of the system, where the fluctuations of $2\left\langle  w_{n}(x) \right\rangle$ become more relevant than in panel \ref{density for L and for R incid aa}b).
Due to this drop, the DMPK result of Eq. \eqref{total density level n}, is not satisfactory to describe the numerical results of $2\left\langle  w_{n}(x) \right\rangle$ for '{\it intermediate}' energy levels.
This fact can be understood from table \ref{Details_wn}, which shows that, for both '{\it intermediate}' energy levels, the system is localized, however, the weak disorder requirement and the random phase approximation are being left. 
%


The numerical results of panels \ref{density for L and for R incid aa}e) and \ref{density for L and for R incid aa}f) show the profiles of $2\left\langle  w_{n}(x) \right\rangle$ for two low-lying energy levels; the energy levels are $n=1500$ and $n= 10$, respectively. 
For those low-lying energy levels, the individual contributions to the electron density, for left incidence $\left\langle w_{n}^{LI}(x) \right\rangle$, and right incidence $\left\langle  w_{n}^{RI}(x) \right\rangle$, banish near to the borders of the disordered system. This means that, for both incidences, the electrons do not penetrate the sample, giving rise to a dramatic drop of the profiles $2\left\langle w_{n}(x) \right\rangle$ inside the system;
therefore, the DMPK result, Eq. \eqref{total density level n}, is not appropriate to describe  $2\left\langle w_{n}(x) \right\rangle$ for low-lying energy levels.
The failure of the DMPK model can be understood from the details shown in table \ref{Details_wn}: the weak disorder condition and the random phase approximation are not satisfied. In panel \ref{density for L and for R incid aa}e), the parameter $z_{n}=k_n \ell_n = 0.23$, while for panel \ref{density for L and for R incid aa}f), $z_{n}=k_n \ell_n = 6 \times 10^{-5}$, which is even farther from the weak disorder requirement. 
In the case of panel \ref{density for L and for R incid aa}f), we have for one scatterer $\langle R_1(k_n) \rangle \approx 0.96$, while for the total sample, $\langle R(k_n) \rangle = 1.000$ and $\langle r(k_n) \rangle = -0.995 - i 0.003$. This does not indicate that the system has become more localized, but rather that each $\delta$ potential strength $u_{j}\gg k_{n}$, Eq. \eqref{Def_Coef_R_1}, while the total sample has becomes more impenetrable.

\subsubsection{Transmission Spectrum}
\label{Results_1_Realization}

The numerical results shown in Figs. \ref{density for L and for R incid c 1} and \ref{density for L and for R incid aa} for $\left\langle w_{n} \left( x \right) \right\rangle $, and its corresponding left $\left\langle w_{n}^{LI} \left( x \right) \right\rangle $ and right $\left\langle w_{n}^{RI} \left( x \right) \right\rangle $ contributions, exhibit that as we go down in energy, the wave function penetrates ever less inside the disordered sample; therefore, the electron is reflected back and the transmission gradually decreases. This fact is illustrated in Fig. \ref{fig:Transmission-spectrum}, where we present the transmission spectra for two different disorder configurations and the average over the ensemble of the transmission spectra.

Panels \ref{fig:Transmission-spectrum}a) and \ref{fig:Transmission-spectrum}b) show the transmission spectra for two different disorder configurations, which differ drastically from each other; as it is expected, each disorder realization has its own resonances for different energy levels. In both cases, the most important resonances are found for high-lying levels $n\gg 1$.
In contrast, there are no resonances for low-lying levels $n \sim 1$, which probably are exponentially narrow, so they cannot be excited at low energy levels. The insets of Figs. \ref{fig:Transmission-spectrum}a) and \ref{fig:Transmission-spectrum}b) show a zoom in on small resonances for {\it intermediate} energy levels. Those transmission resonances are low in absolute terms, but they are relatively high, compared to the transmission of their neighbors; a Lorentzian distribution was used to fit the data of those resonances showing an excellent agreement. 
The numerical evidence of panels \ref{fig:Transmission-spectrum}a) and \ref{fig:Transmission-spectrum}b) means that, for a given disorder configuration, the transmission 
coefficient $T\left(\epsilon_{n} \right) $ is only relevant for high-lying levels,
while intermediate levels show very small resonances and the resonances of low-lying levels  are not excited.

\begin{figure}[t]
\includegraphics[width=\columnwidth]{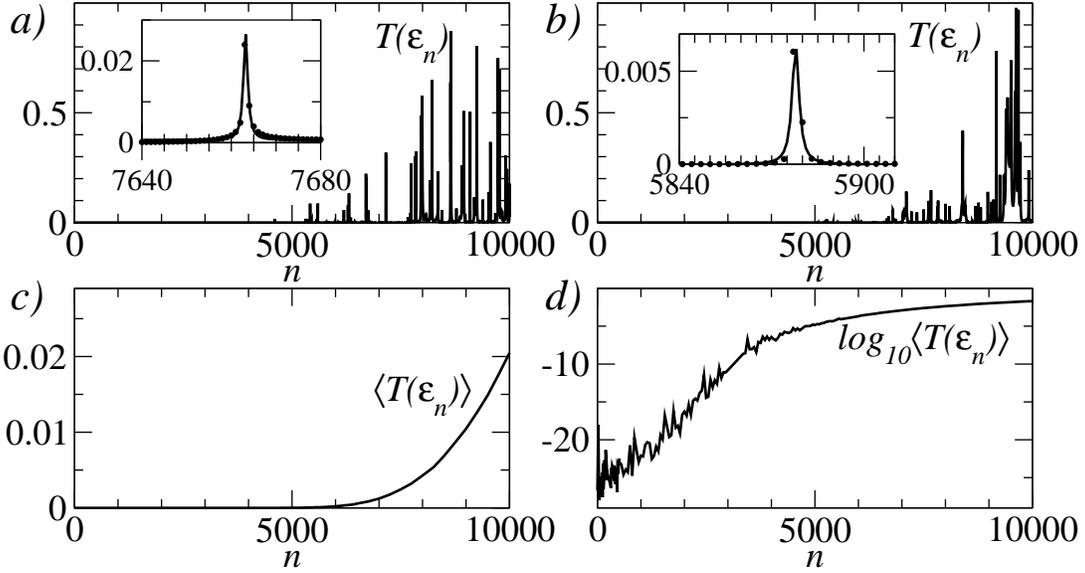}
\caption{Panels a) and b) show transmission spectra $T\left( \epsilon_{n}  \right) $ $vs$ $n$ for two different disorder realizations:
insets show Lorentzian distribution fits for resonances of intermediate energy levels. c) Average transmission spectrum $\left\langle T\left(\epsilon_{n}  \right)  \right\rangle$ d) Logarithm of the average transmission spectrum $\log _{10}\left\langle T\left(\epsilon_{n}  \right)  \right\rangle$.
}
\label{fig:Transmission-spectrum}
\end{figure}

Panel \ref{fig:Transmission-spectrum}c) shows the ensemble average of the transmission spectra $ \left\langle T\left(\epsilon _{n} \right) \right\rangle $; the behavior is in good agreement with those results of Figs. \ref{density for L and for R incid c 1} and \ref{density for L and for R incid aa}, i.e., in average, the transmission decreases as we go down the energy spectrum. 
Finally, panel \ref{fig:Transmission-spectrum}d) shows $\log_{10}\left\langle T\left(\epsilon _{n} \right) \right\rangle $, which emphasizes the drop of the average transmission when the energy decreases.

\subsubsection{The Local Density of States and the Dwell Time}
\label{LDOS_dwell_Time}

We now offer a complementary interpretation of those results of Sec. \ref{density_wn_Sim_DMPK}. The discussion is based on the {\it local density of states} (LDOS) $\bm\rho_{_{LDOS}}\left(\epsilon_{n},x\right)$ and the {\it dwell time} $\tau_{_{D}}\left( \epsilon_{n}\right)$: see App. \ref{DOS and Tau}.

The LDOS is the sum over left and right incidences of particle densities with unitary flux; therefore, for the $n$-th level, the averages $\left\langle w_{n}\left(x\right) \right\rangle $, Eqs. \eqref{Individual_level_w_n}-\eqref{wn_LI_RI}, and $\left\langle \bm\rho_{_{LDOS}}\left(\epsilon_{n},x\right)\right\rangle $, Eqs. \eqref{Def_LDOS_1}-\eqref{Def_LDOS_2}, are related in the following way:
\begin{equation}
2 \pi \hslash v_{n} \left\langle \bm\rho_{_{LDOS}}\left(\epsilon_{n},x\right) \right\rangle 
= 2
\left\langle w_{n}\left( x \right) \right\rangle
=
\left\langle w^{LI}_{n}\left( x \right) \right\rangle+\left\langle w^{RI}_{n}\left( x \right) \right\rangle
;
\label{Aver_LDOS_wn_vn}
\end{equation}
here, $v_{n} \equiv \hslash k_{n}/m$ denotes the unitary flux for the $n$-th energy level, which coincides with the group velocity defined in Eq. \eqref{density of states c}. Equation \eqref{Aver_LDOS_wn_vn}, relates the average LDOS $\left\langle \bm\rho_{_{LDOS}}\left(\epsilon_{n},x\right) \right\rangle $ to the results found in Sec. \ref{density_wn_Sim_DMPK} for $\left\langle w_{n}\left( x \right) \right\rangle$; we focused the analysis on the region inside the sample, i.e., $0\leq x \leq L$.

The numerical results shown in Figs. \ref{density for L and for R incid c 1}, \ref{density for L and for R incid aa}a) and \ref{density for L and for R incid aa}b), corresponding to high-lying energy levels, satisfy the DMPK prediction of Eq. \eqref{total density level n}. That is,
\begin{equation}
\left\langle w_{n}\left( x \right) \right\rangle = \left\langle \mathcal{U}\left(\epsilon_{n},x\right) \right\rangle 
\frac{v_{n}}{2}
= 1, \;\; n \gg 1,
\label{Aver_LDOS_n gg_1}
\end{equation}
where we have defined the following quantity,
\begin{equation}
\left\langle \mathcal{U}\left(\epsilon_{n},x\right) \right\rangle  \equiv 2\pi \hslash \left\langle \bm\rho_{_{LDOS}}\left(\epsilon_{n},x\right) \right\rangle .
\label{Huang_Quantity}
\end{equation}
The result given in Eq. \eqref{Aver_LDOS_n gg_1} for high-lying energy levels, agrees with the experimental results of a recent microwave measurement of energy density inside lossless 1D random media studied by Huang, \textit{et. al.} \cite{Huang_2022}; the experimental setup was a single-mode random waveguide of copper with cutoff frequency of $6.56$ $\mathrm{GHz}$. The experimental results were presented for the electromagnetic version of Eq. \eqref{Huang_Quantity}, i.e., 
\begin{equation}
\left\langle \mathcal{U}^{Elec}\left(\omega,x\right) \right\rangle \equiv 2\pi \left\langle \bm\rho_{_{LDOS}}^{Elec}\left(\omega,x\right) \right\rangle,
\label{Huang_Quantity_bis}
\end{equation}
which satisfies Eq. \eqref{Aver_LDOS_n gg_1}: see Fig. $2$ of Ref.  \cite{Huang_2022}. The measurements were made in the frequency range $10.00-10.70$ $\mathrm{GHz}$, with disordered samples of length $L=86$ $\mathrm{cm}$ and a ratio $L/\ell=3.51$, $L$ being the system length and $\ell$ the mean free path; therefore, Huang's experiment satisfies the weak disorder condition $k\ell\simeq 51.3 \gg 1$, as those results shown in Figs. \ref{density for L and for R incid c 1}, \ref{density for L and for R incid aa}a) and \ref{density for L and for R incid aa}b), where $k_{n}\ell_{n}= 55.04$, $61.6$ and $21.7 $, respectively: see table \ref{Details_wn}.

The numerical results shown in Fig. \ref{density for L and for R incid aa} exhibit that, as we go down the spectrum, the average dimensionless density $2 \left\langle w_{n}\left( x \right) \right\rangle$ drops at the center of the sample. From Eq. \eqref{Aver_LDOS_wn_vn}, this means that the average LDOS $\left\langle \bm\rho_{_{LDOS}}\left(\epsilon_{n},x\right) \right\rangle $ is depleted in the interior of the system as the energy level decreases. 
In order to understand this, we present in Fig. \ref{fig:w_individual_realization}, for a given disorder realization, some relevant profiles of the dimensionless densities for left $w_{n}^{LI}\left(x\right) $ and right $w_{n}^{RI}\left(x\right) $ incidences; the profiles correspond to the transmission spectrum of panel \ref{fig:Transmission-spectrum}b).

\begin{figure}[t]
\includegraphics[width=\columnwidth]{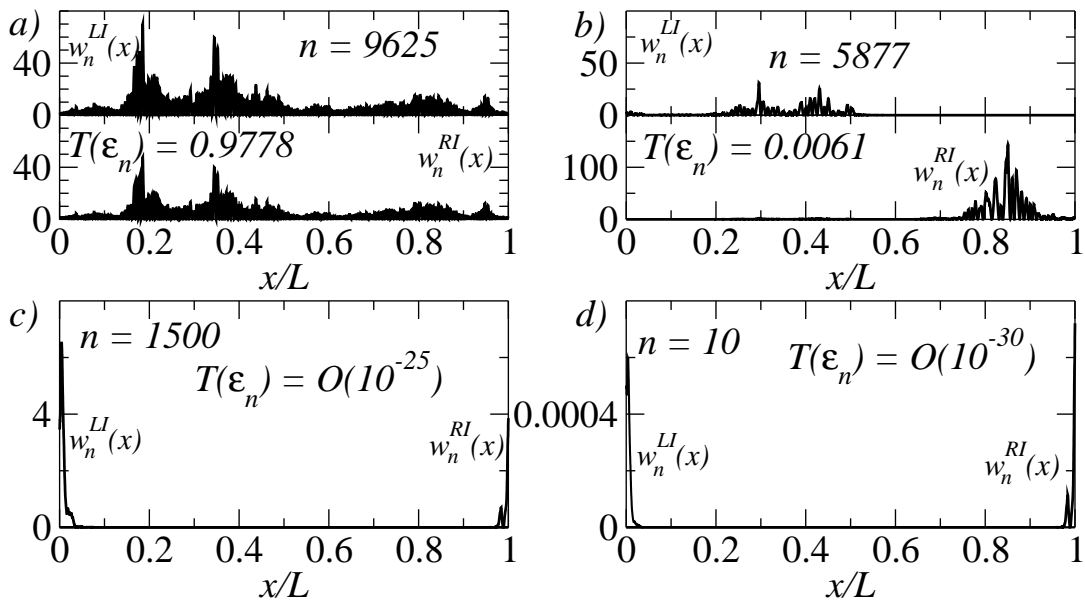}
\caption{This figure shows, for a particular disorder realization, the dimensionless electron density profile inside
the disordered sample for left, $w_{n}^{LI}\left(x\right)$, and right,
$w_{n}^{RI}\left(x\right)$, incidence for four different energy levels. Panels a) and b) correspond to resonant levels, and are divided in two plots, one on top
of the other, to better visualize the behavior of each incidence.
Panels c) and d) correspond to non-resonant levels.
}
\label{fig:w_individual_realization}
\end{figure}

Panel \ref{fig:w_individual_realization}a) is divided in two plots: the upper panel for $w_{n}^{LI}\left( x\right)$ and the lower panel for $w_{n}^{RI}\left( x\right)$. These profiles correspond to a high-lying energy level, where the highest resonance of the spectrum of panel \ref{fig:Transmission-spectrum}b) takes place. We observe that, for both incidences, the electron density profile is extended inside the sample; therefore, the wave function penetrates the sample, giving rise to a relevant transmission coefficient $T\left(\epsilon_{n}\right)\sim 1$.

Panel \ref{fig:w_individual_realization}b) is also divided in two plots: the upper panel for $w_{n}^{LI}\left( x\right)$ and the lower panel for $w_{n}^{RI}\left( x\right)$; these profiles correspond to an intermediate low resonant energy level [the one shown in the inset of Fig.\ref{fig:Transmission-spectrum}b)]. In this case, both incidences $w_{n}^{LI}\left( x\right)$ and $w_{n}^{RI}\left( x\right)$, show that the wave function is not extended inside the system, in accordance with a low transmission coefficient  $T\left(\epsilon_{n}\right)\ll 1$; however, the electron penetrates a short distance into the sample, and then it seems to be localized in a very narrow region inside the sample.

In panels \ref{fig:w_individual_realization}c) and \ref{fig:w_individual_realization}d), the behavior of $w_{n}^{LI}\left( x\right)$ and $w_{n}^{RI}\left( x\right)$ is quite notorious. These profiles of low-lying levels, show a dramatic drop near the borders of the sample, so the wave function does not penetrate the disordered region and the transmission coefficient is negligible  $T\left(\epsilon_{n}\right)\to 0$.

Roughly speaking, the results of Fig. \ref{fig:w_individual_realization} show that, for a given disorder configuration, the electron propagates through the sample '{\it spending}' time according to the energy level $n$ of the incident wave function. 
This qualitative interpretation can be analyzed by using the concept of '{\it dwell time}'. The dwell time is a measure of the time spent by an electron in the disordered region $0 < x < L$ regardless of whether it is ultimately transmitted or reflected \cite{winful_delay_2003}.

From App. \ref{DOS and Tau}, the averages of the dwell time, the LDOS and the dimensionless density are related in the following way
\begin{equation}
\left\langle \tau_{_{D}}\left( \epsilon_{n}  \right)  \right\rangle
\equiv 2\pi \hslash 
\frac{1}{v_{n}}\int_{0}^{L} 
\left\langle \bm\rho_{_{LDOS}}\left(\epsilon_{n},x\right) \right\rangle dx
 = \frac{1}{v_{n}}\int_{0}^{L}
2 \left\langle  w_{n}\left( x\right) \right\rangle  dx.
\label{Aver_Dwell_w_n}
\end{equation}

Figure \ref{fig:dwell time plot} shows the numerical results for the average dwell time $\left\langle \tau_{_{D}}\left(\epsilon_{n}  \right) \right\rangle/\tau_{_{0}}$ in units of the characteristic time
\begin{equation}
\tau_{_{0}}=\frac{L}{v_{0}};
\end{equation}
here $L$ is the system length, while 
\begin{equation}
v_{0}=\frac{\hslash u_{0}}{m},
\end{equation}
is a characteristic velocity defined in terms of the Planck's constant $\hslash$, the electron mass $m$ and the maximum value of the delta potential strength $u_{0}$: see Eq. \eqref{MicrocopicPotentialModel}. 
We notice that, in average, the dwell time goes to zero in the strong scattering regime, i.e., for low-lying energy levels $n \sim 1$, where the strength of the scatterers is much larger than the energy; in this regime, the wave function does not penetrate the sample, the electron is reflected back, the LDOS is depleted and the dwell time is too short. In the weak disorder regime, for high-lying energy levels $n\gg 1$, we observe a smooth decay that do not reach zero; in this case,
$\left\langle \bm\rho_{_{LDOS}}\left(\epsilon_{n},x\right) \right\rangle$ satisfies Eq. \eqref{Aver_LDOS_n gg_1}, the wave function penetrates the system and the electron propagates through the sample with a short dwell time. The most notorious behavior shown in Fig. \ref{fig:dwell time plot} is for intermediate levels: as the whole sample becomes more and more impenetrable the transmission is negligible, the average dimensionless density  $\left\langle   w_{n}\left( x\right) \right\rangle  $ drops and the average dwell is long for intermediate energy levels but decreases for both, low-lying levels and high-lying levels.

The behavior shown for $\left\langle \tau_{_{D}}\left( \epsilon_{n}  \right)  \right\rangle $ in Fig. \ref{fig:dwell time plot}, is similar, at least qualitatively, to the results of the dwell time in potential barriers \cite{buttiker_larmor_nodate, winful_delay_2003}.

\begin{figure}[t]
\centerline{
\includegraphics[width=9cm,height=7cm]{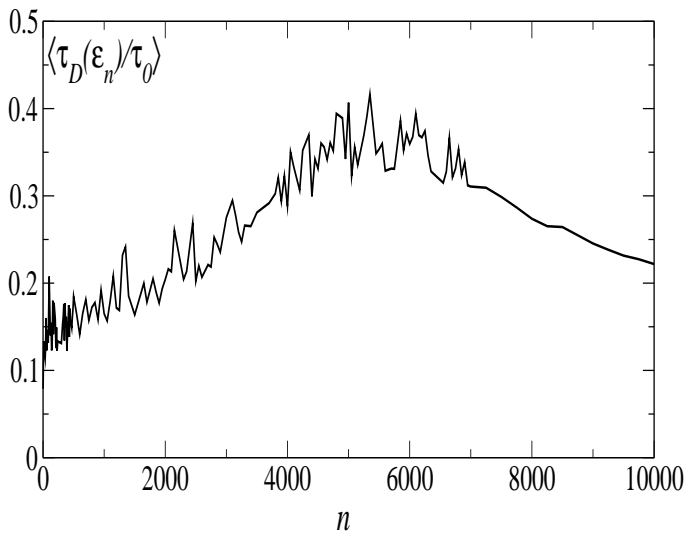}
}
\caption{Average dwell time as a function of the energy level. The characteristic time $\tau_{0}=mL/\hslash u_{0}$, where $u_{0}$ is the maximum value of the delta potential strength.
}
\label{fig:dwell time plot}
\end{figure}

\subsubsection{Electromagnetic analogue}

The DMPK prediction, Eq. \eqref{total density level n}, and our numerical simulations for high-lying energy levels shown in Sec. \ref{density_wn_Sim_DMPK}, agrees with recent experimental results of microwave measurement in a single-mode random waveguide \cite{Huang_2022}.
This is because the propagation of electromagnetic waves through a waveguide is the optical analogue of electron conduction through a wire  \cite{Beenakker_1997}. 
The analogy is direct if we simplify the vector nature of the electromagnetic waves to a scalar description. This is done by considering a two-dimensional waveguide of width $W$ with a Transverse Electric Wave; in this case, the electric field $ \overrightarrow{ \mathcal{E}}\left( x,y\right) = E\left(x,y\right)e^{-i\omega t} \widehat{\bm z} $ (frequency $\omega$ and wavenumber $k=\omega/c$), has a perpendicular polarization to the propagation direction $x$. The scalar complex field $E\left(x,y\right)$ satisfies the scalar Helmholtz equation
\begin{equation}
\left[ \nabla^{2}+k^{2}\varepsilon\left(x,y\right) \right] E\left(x,y\right)=0,
\end{equation}
with boundary conditions $E=0$. The relative dielectric constant $\varepsilon\left(x,y\right)=1+\delta\varepsilon\left(x,y\right)$ fluctuates due to disorder in the waveguide. If the waveguide supports only one propagating mode, then this electromagnetic analogue can be described as one-dimensional problem, similarly to the 1D stationary Schr\"odinger equation.

Although it is possible to extend some predictions of electron or quantum waves to classical waves, it is important to notice that there exist relevant differences. For instance, in a single mode electromagnetic waveguide, as we go down the frequency, the propagating mode would be closed, so we have no wave propagation; in addition, the electron scattering increases as the energy decreases, but the scattering of light falls at low frequencies. Due to these facts, it is complicated to compare our numerical results with microwave experiments in a single-mode waveguide.

\subsubsection{Summary}


The results presented above allow us to conclude that, as we go down in energy, the wave function penetrates ever less inside the sample.
The interpretation of $s_n$ in terms of a mfp due to localization from disorder gradually gives way to an interpretation as a parameter that measures the extent to which 
the wave function is reflected back
because each scatterer is 'seen' by the electron as a higher and higher 
--and hence impenetrable-- 
potential barrier.

The following comments are in order. 
The results in panel a) of Fig. \ref{density for L and for R incid aa} 
are of the same nature as those of Ref. \cite{mello-shi-genack} and Ref. \cite{Huang_2022}.
As we go down the spectrum with $N_{scatt}$ {\it fixed} ($L$ and $d$ 
{\it fixed}), 
i.e., to panel b) and all the way to f),
we have the behavior which was described above.

In contrast, in Ref. \cite{mello-shi-genack}  $s_n$ was increased keeping 
{\it ${\ell}_n$ fixed} and increasing $N_{scatt}$  (with $d$ {\it fixed}, so $L$ 
{\it increases}).
The result is a  legitimate increase in localization.
I.e.,  
starting from a density profile of a similar nature as that of panel a) of 
Fig. \ref{density for L and for R incid aa} of the present paper, 
the result of increasing $s_n$ was that 
the density profile for left incidence approached 2 in the left half of the sample, and 0 in the right half.
The behavior of LDOS is just the same.

LDOS is sensitive to these two different procedures,
while $s_n$ itself is not.

We mention that the numerical results of Fig. \ref{density for L and for R incid aa} show that, for a rather wide range of $n \gg 1$, the sum of both incidences is $x$-independent, as predicted by Eq. \eqref{total density level n}.
Moreover, for levels near the Fermi level, the numerical results for left 
$\left\langle w_{n}^{LI}(x)\right\rangle$ and right $\left\langle w_{n}^{RI}(x)\right\rangle$ 
incidence are in good agreement with the DMPK prediction found in 
Ref. \cite{mello-shi-genack}. 

As the energy level decreases, the numerical results exhibit quite important statistical fluctuations at the center of the system, which is an effect of the finite size of the sample ($10^{6}$ realizations): see panels c) and d).
Fluctuations in Fig. \ref{density for L and for R incid aa} appear when we go down the energy level, and thus the transmission decreases. A similar behavior occurs in multichannel systems: low transmission eigenchannels present higher fluctuations, compared to high transmission eigenchannels \cite{bender2020fluctuations}, and are strongly correlated over a wide energy range \cite{shi2015dynamic}; also, as a consequence of spectral correlations, long-range correlations are expected to be observed in long  samples \cite{genack1990long}.

The results of Sec. \ref{density_wn_Sim_DMPK} show that, the wave function penetrates ever less inside the sample. This is reflected in the fact that the LDOS is depleted in the interior of the system, since the wave function is ever smaller inside. This result was related to the dwell time, which allows us to interpret the depleted of the LDOS.

\subsection{All levels}
\label{density W}

We now analyze the expectation value of the {\it total density}, 
$\left\langle {\cal W}\left( x\right)/{\cal W}_{0}\right\rangle $. 
From Eqs. \eqref{a} and \eqref{total density level n}, the theoretical  DMPK prediction is 
\begin{equation}
\left\langle \frac{{\cal W}(x )}{{\cal W}_0} \right\rangle_{DMPK}
=1,\;\;\; \forall x,
\label{W/W0_DMPK_inside_outside}
\end{equation}
which is, once again, insensitive to disorder. 
Fig. \ref{density for L and for R incid and all levels 1} compares
the result of Eq. \eqref{W/W0_DMPK_inside_outside} with computer simulations.
Inside the disordered region, DMPK gives an unsuitable description.
The discrepancy is due to the low-lying levels that behave very differently from the high-lying ones, as explained in the text in relation with Fig. \ref{density for L and for R incid aa}.

Fig. \ref{density for L and for R incid and all levels 1} also shows statistical fluctuations, due to the finite size of the sample ($10^{5}$ realizations). Outside the disordered system, the theoretical prediction of the DMPK model is in good agreement with numerical results, except in the vicinity of the left reservoir, $x \simeq-L_{0}/2$, where the numerical result shows a minimum, while DMPK does not. The asymmetry in the numerical result with respect to the right reservoir, which does not show a minimum, is due to the off-center position of the sample, which lies in the region $x \in [0,L]$. Should the sample be centered, i.e., in $x \in [-L/2, L/2]$, such asymmetry in the numerical results would not be there.

\begin{figure}[h]
\includegraphics[width=12cm,height=8cm]{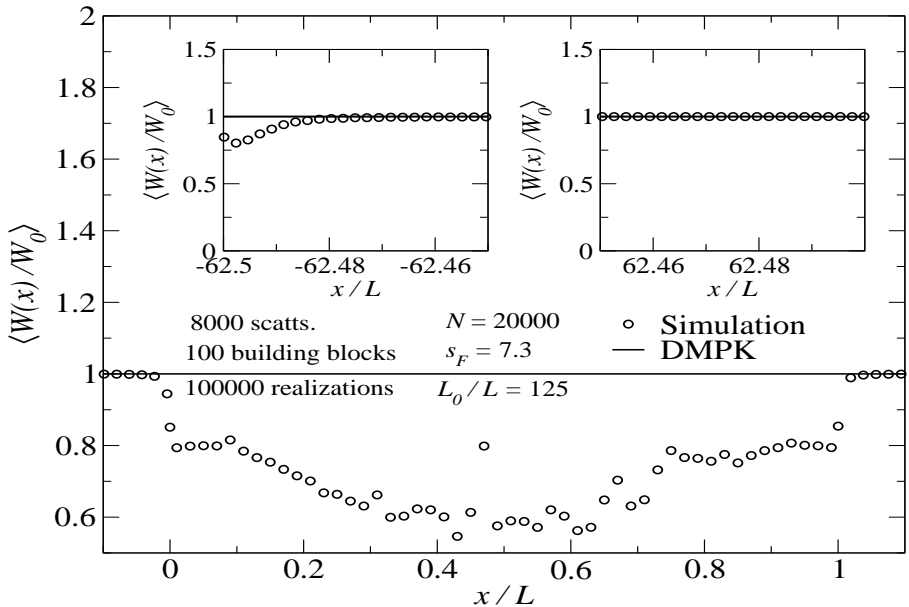}
\caption{
Numerical results for $\left\langle \frac{{\cal W}(x)}{{\cal W}_0} \right\rangle$ {\it vs} $x/L$ for a disordered system of length
$L$ in {\it equilibrium} at $T=0$, with the same chemical potential at both reservoirs. The total system, consisting of the sample 
plus the ballistic regions, 
has a length $L_0/L = 125$. We are considering $N= 20000$ states, i.e.,  $10,000$ levels, each level consisting of two states, 
one traveling to the left and one to the right. $N$ is thus the total number of electrons. The quantity $s_F=L/\ell_F= 7.3$ for the Fermi level, and increases as the 
level goes down toward the ground state. The simulation consists of $10^5$ realizations of disordered systems, each one with $8000$ 
scatterers. To reduce the computing time, the simulation generates every disordered system by combining $100$ building blocks, each containing $80$ individual scatterers.
}
\label{density for L and for R incid and all levels 1}
\end{figure}

\section{The expectation value of the logarithm of the density in equilibrium at zero temperature}
\label{aver log density}

\subsection{Individual levels}
\label{Individual levels log}
\subsubsection{Outside the sample}

Consider first $x<0$.
For one level $n$, with left and right incidence, Eq. (\ref{wn x<0}) gives
\begin{subequations}
\begin{eqnarray}
\left\langle\ln w_n(x<0)\right\rangle
&=& \left\langle
\ln \left\{
1+ \frac12 \left[r(k_n)e^{-2ik_n x}+ r^{*}(k_n)e^{2ik_n x}\right]
\right\}  
\right\rangle
\\
&=& \Big\langle
\ln \Big[
1- |r(k_n)| \cos2(
\nu\left( k_{n}\right) 
- k_nx)
\Big] 
\Big\rangle \\
&=& \int_0^{\infty}d\lambda \int_0^{2\pi} d \nu \;
p_{s_{n}}(\lambda, \nu)
\ln \Big[
1- \sqrt{\frac{\lambda}{1+\lambda}}\cos2(\nu - k_nx)
\Big],
\end{eqnarray}
\end{subequations}
where we have used the polar representation of the reflection amplitude, $r = \sqrt{\lambda/(1+\lambda)}e^{2i\nu}$ \cite{mello-kumar}, and the energy dependence has been obviated. Let
\begin{eqnarray}
d\lambda \; d \nu \; p_{s_{n}}(\lambda, \nu)
&=& d\lambda \; p_{s_{n}}(\lambda) \; \frac{d \nu}{2 \pi}
\end{eqnarray}
Then
\begin{eqnarray}
\left\langle\ln w_n(x<0)\right\rangle
&=& \int_0^{\infty}d\lambda \; p_{s_{n}}(\lambda)
I_n(\lambda)
\label{Av ln wn out L}
\end{eqnarray}
where
\begin{subequations}
\begin{eqnarray}
I_n(\lambda)
&=& \int_0^{2\pi} \frac{d\nu}{2\pi}
\ln \Big[
1- \sqrt{\frac{\lambda}{1+\lambda}}\cos2(\nu - k_nx)
\Big]  \\
&=&  \frac{1}{\pi}\int_0^{\pi} 
\ln \left[1- \sqrt{\frac{\lambda}{1+\lambda}} \cos \phi \right]  d \phi  \\
&=& \ln \frac{\sqrt{1 + \lambda}+1}{2\sqrt{1+\lambda}}
\end{eqnarray}
\label{av ln dens left arm}
\end{subequations}

Consider now $x > L$. 
For one level $n$, with left and right incidence, Eq. (\ref{wn x>L}) gives
\begin{subequations}
\begin{eqnarray}
\left\langle\ln w_n(x > L)\right\rangle
&=& \left\langle
\ln \left\{
1+ \frac12 \left[r'(k_n)e^{2ik_n x}+ (r'(k_n))^{*}e^{-2ik_n x}\right]
\right\}  
\right\rangle
\\
&=& \Big\langle
\ln \Big[
1+ |r'(k_n)| \cos2(
\mu\left(k_{n}\right)
+ k_nx)
\Big] 
\Big\rangle  \\
&=& \int_0^{\infty}d\lambda \int_0^{2\pi} d \mu \;
p_{s_{n}}(\lambda, \mu)
\ln \Big[
1+\sqrt{\frac{\lambda}{1+\lambda}}\cos2(\mu + k_n x)
\Big] 
\end{eqnarray}
\label{Av ln wn out}
\end{subequations}
Let
\begin{eqnarray}
d\lambda \; d \mu \; p_{s_{n}}(\lambda, \mu)
&=& d\lambda \; p_{s_{n}}(\lambda) \; \frac{d \mu}{2 \pi}
\end{eqnarray}
Then
\begin{eqnarray}
\left\langle\ln w_n(x>0)\right\rangle
&=& \int_0^{\infty}d\lambda \; p_{s_{n}}(\lambda)
J_n(\lambda)
\label{Av ln wn out}
\end{eqnarray}
where
\begin{subequations}
\begin{eqnarray}
J_n(\lambda)
&=& \int_0^{2\pi} \frac{d\mu}{2\pi}
\ln \Big[
1+ \sqrt{\frac{\lambda}{1+\lambda}}\cos2(\mu + k_nx)
\Big]  \\
&=&  \frac{1}{\pi}\int_0^{\pi} 
\ln \left[1+ \sqrt{\frac{\lambda}{1+\lambda}} \cos \phi \right]  d \phi
\\
&=& \ln \frac{\sqrt{1 + \lambda}+1}{2\sqrt{1+\lambda}}
\end{eqnarray}
\end{subequations}

\subsubsection{Inside the sample: $x\in [0,L]$}

From Eq. \eqref{mn(x)inside}
\begin{eqnarray}
 \Big\langle  \ln {w}_n(x \in [0,L])  \Big\rangle
&=& \left\langle 
\ln{\frac12}
\Big[
F_{x}\left( M_2 \left(\epsilon_{n}\right) \right) + G_{x}\left( M_1 \left(\epsilon_{n}\right) \right)
\Big]      \right\rangle
+\langle  \ln T\left( \epsilon_{n} \right)  \rangle,
\label{Av ln wn inside}
\end{eqnarray}
where $\left\langle \ln T\left( \epsilon_{n} \right) \right\rangle =-L/\ell_{n} $ (see Ref.  \cite{mello-kumar}). In Eq. \eqref{Av ln wn inside}, we have defined the following expressions
\begin{subequations}
\begin{eqnarray}
F_{x}\left( M_2 \left(\epsilon_{n}\right) \right)
&\equiv&
\Big| 
\alpha_2^{*}(\epsilon_n){\rm e}^{ik_{n}x} - \beta_2^{*}(\epsilon_n){\rm e}^{-ik_{n}x}
\Big|^2 ,
\\
G_{x}\left( M_1 \left(\epsilon_{n}\right) \right)
&\equiv&
\Big| 
\beta_1(\epsilon_n){\rm e}^{ik_{n}x} +\alpha_1^{*}(\epsilon_n){\rm e}^{-ik_{n}x}
\Big|^2 ,
\end{eqnarray}
\end{subequations}
\noindent where the disorder sample has been divided in two subsamples, one to the left of the point $x$, and the other to the right; each subsample has its own transfer matrix: $M_1$ for the left subsample, and $M_2$ for the right subsample. See App.\ref{proof eq Wn+}. So, using the polar representation for $M_1$ and $M_2$, we have
\begin{eqnarray}
\ln{\frac12}  
\Big[
F_{x}\left( M_2 \left(\epsilon_{n}\right) \right) + G_{x}\left( M_1 \left(\epsilon_{n}\right) \right)
\Big]   
&=&\ln \Big[
C \Big( \lambda_1\left( k_{n} \right), \lambda_2 \left( k_{n} \right) \Big) 
\nonumber
\\
&-& D \Big(  \lambda_2 \left( k_{n} \right) \Big) 
\cos [2(k_{n}x - \nu_2 \left( k_{n} \right)) ]
\nonumber
\\
&+& D \Big(  \lambda_1 \left( k_{n} \right) \Big) 
\cos[2(k_{n}x+ \mu_1 \left( k_{n} \right))]
\Big]
\end{eqnarray}
\begin{figure}[b]
\includegraphics[width=10cm,height=6cm]{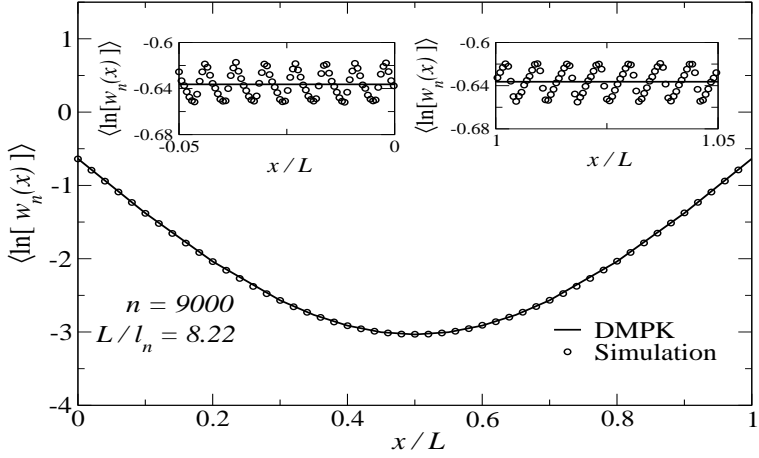}
\caption{The quantity $\left\langle \ln w_n(x) \right\rangle$ 
for the $n=9000$ level
from computer simulations, compared with 
the DMPK prediction, Eq. (\ref{Av ln wn inside}),  (\ref{Av ln wn in}), inside the sample and 
(\ref{Av ln wn out L}), (\ref{Av ln wn out}) outside the sample.
The system is in equilibrium at $T=0$, with the same chemical potential at both reservoirs.
The insets represent in more detail the results outside the system proper.
We have a sample with $10^6$ realizations.
The agreement between theory and numerical simulations is excellent.
}
\label{aver ln W level n}
\end{figure}
with
\begin{subequations}
\begin{eqnarray}
C \Big( \lambda_1\left( k_{n} \right), \lambda_2 \left( k_{n} \right) \Big) 
&=&   1 + \lambda_1 \left( k_{n} \right) + \lambda_2 \left( k_{n} \right) \equiv C
\\
D \Big(  \lambda_1 \left( k_{n} \right) \Big) 
&=&   \sqrt{\lambda_1 \left( k_{n} \right)  (1+ \lambda_1 \left( k_{n} \right))} \equiv D_1        \\
D \Big(  \lambda_2 \left( k_{n} \right) \Big) 
&=&\sqrt{\lambda_2 \left( k_{n} \right) (1+ \lambda_2 \left( k_{n} \right))} \equiv D_2
\end{eqnarray}
\end{subequations}
and its average
\begin{eqnarray}
&&\left\langle \ln{\frac12}  \Big[F_x (M_2 \left( k_{n} \right) ) + G_x(M_1 \left( k_{n} \right) ) \Big]   \right\rangle
= \int_{0}^{\infty}  \int_{0}^{\infty} d\lambda_1 d\lambda_2
p_{s^{(1)}_{n}}(\lambda_1)p_{s^{(2)}_{n}}(\lambda_2)
\nonumber 
\\
&&\times \int_0^{2\pi}\frac{d\mu_1}{2\pi} \int_0^{2\pi}\frac{d\nu_2}{2\pi}
\ln \Big[
C + D_1 \cos 2 (k_{n}x + \mu_1) - D_2 \cos 2(k_{n}x - \nu_2)
\Big],
\label{Av ln wn in} 
\nonumber  \\ 
\end{eqnarray}
once again, we omitted the energy dependence of the polar parameters. One of the angular integrations gives
\begin{subequations}
\begin{eqnarray}
&&\int_0^{2\pi}\frac{d\mu_1}{2\pi} \int_0^{2\pi}\frac{d\nu_2}{2\pi}
\ln \Big[
C + D_1 \cos 2 (k_{n}x + \mu_1) - D_2 \cos 2(k_{n}x - \nu_2)
\Big] \\
&=& \frac{1}{\pi}
\int_0^{\pi}  d\phi
\ln \left\{
\frac12 \Big[
(C+D_1 \cos \phi) + \sqrt{(C+D_1 \cos \phi)^2 - D_2^2}
\Big]
\right\}
\nonumber \\
\end{eqnarray}
\end{subequations}
This last result has to be inserted in (\ref{Av ln wn in}) and this in (\ref{Av ln wn inside}).
The $x$ dependence of the result appears in the fact that  
$s^{(1)}_{n}=x/\ell_n$ and $s^{(2)}_{n}=(L-x)/\ell_{n}$, which
denote, respectively, the scaled 
lengths at left and right of the position $x$.
The problem of finding $\left\langle {w}_n(x)\right\rangle$ is thus reduced to quadratures.

The comparison with computer simulations is given in Fig. \ref{aver ln W level n}}. We observe that the agreement is excellent.

Notice that the average of the logarithm of the electron density inside the sample for a high-lying state would be a decreasing straight line if we had only incidence from the left, just as in 
Ref.  \cite{cheng-yepez-mello-genack}, 
and a symmetrical one for incidence from the right; the combination of the two gives the result of Fig. \ref{aver ln W level n}.

\begin{figure}[b]
\includegraphics[width=10cm,height=8cm]{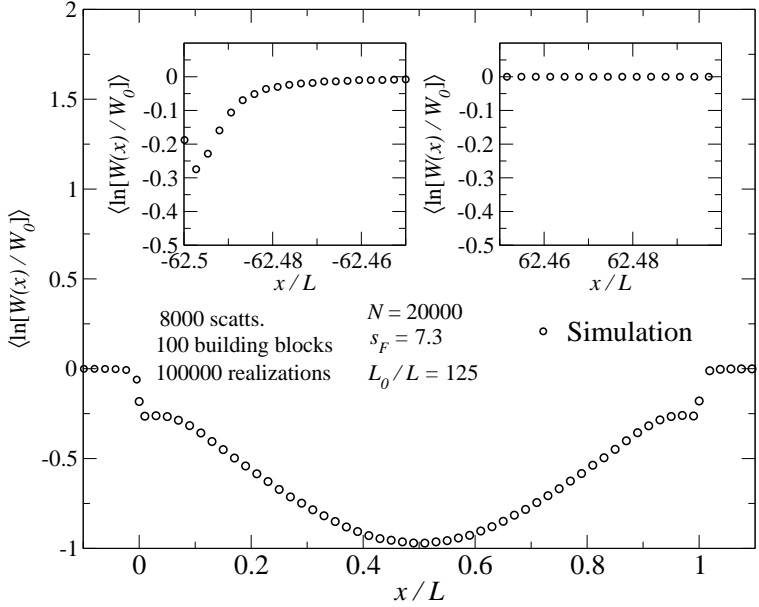}
\caption{
Computer simulation for the expectation value of the logarithm of the density, 
$\left\langle \ln \frac{{\cal W}(x)}{{\cal W}_0} \right\rangle$, outside and inside the sample, whose length is $L$.
The abscissa is $x/L$.
The system is in equilibrium at $T=0$, with the same chemical potential at both reservoirs.
The various parameters used in the simulation are as follows: 
1) the total system, consisting of the sample plus the ballistic regions, has a length 
$L_0/L = 125$;
2) The total number of states is $N=2 \times 10^4$, i.e., $10^4$ levels, each level consisting of two states, one traveling 
to the left and one to the right;
3) $s_F=L/\ell_F= 7.3$ for the Fermi level, and increases as the level goes down toward the ground state;
4) The details of the simulation are the same as those explained in Fig. \ref{density for L and for R incid and all levels 1}.
}
\label{aver ln W all levels}
\end{figure}

\subsection{All levels}
\label{av ln W all levels}

In this case we have not succeeded in finding a theoretical prediction.
Thus, we only present the result of a computer simulation in Fig. \ref{aver ln W all levels}.

\section{The non-equilibrium ($\mu_1>\mu_2$) expectation value of the density and of its logarithm at zero temperature. The contribution of all levels}
\label{density_dmpk_no_equil}

We now consider our system to be at zero temperature, but with a 
{\it non-zero chemical potential difference} between the two reservoirs. 

\begin{figure}[t]
\centerline{
\includegraphics[width=10cm,height=8cm]{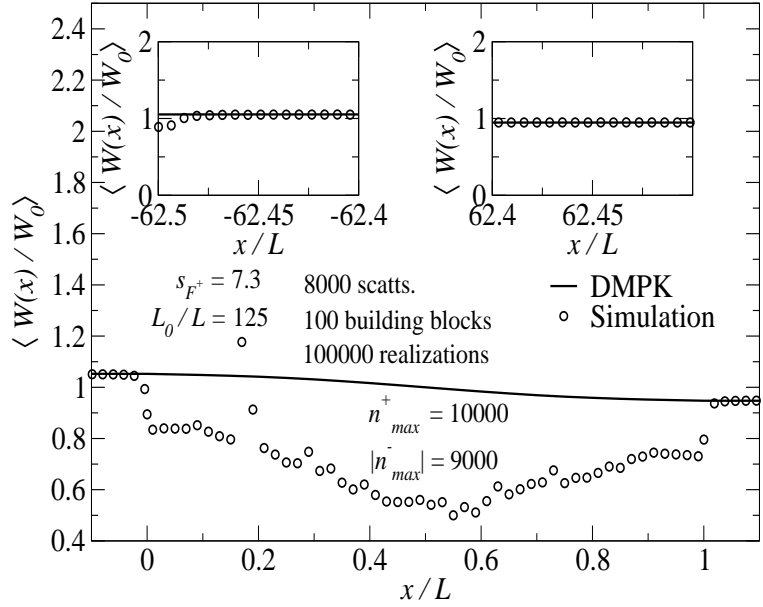}
}
\caption{
The average electron density at zero temperature, when the left reservoir has a higher chemical  potential then the right 
one ($\mu_1 > \mu_2$).
It is expressed in units of its value for a non-disordered system, 
$\langle W(x)/W_0\rangle$. 
The DMPK approximation (continuous line) is compared with a computer simulation (circles).
The insets show the density outside the sample in more detail.
The value $L_0/L= 125$ was used.
The ratio of the sample length to the mfp for the left Fermi level
is $s_{F^{+}}=L/\ell_F = 7.3$;
the ratio of the number of states between the lower and upper chemical potentials $\Delta N=n_{max}^{+}-\vert n_{max}^{-}\vert$ and the total number of states $N$ is
$\Delta N/N=0.053$.
The details of the simulation are the same as those explained in Fig. \ref{density for L and for R incid and all levels 1}.
}
\label{W(x) in DMPK 0}
\end{figure}

Fig. \ref{W(x) in DMPK 0} shows the contribution of all levels to the averaged electron density inside and outside the system, as predicted by the {\it DMPK model}, Eqs. (\ref{<W(x)> outside in DMPK 1}) and  (\ref{<W> inside, DMPK 1}), as well as a computer simulation. The agreement between the two descriptions in the left and right ballistic regions is very good, while the discrepancy inside the disordered system has the same origin as in the previous figures; the DMPK prediction, Eq. \eqref{<W> inside, DMPK 1 d}, gives a small correction over $\left\langle W(x)/{\cal W}_0 \right\rangle  = 1$, which is positive for $x\in (0,L/2)$ and negative for $x\in (L/2,L)$, while the numerical simulations show a deep fall inside the disordered system.

\begin{figure}[h]
\centerline{
\includegraphics[width=10cm,height=8cm]{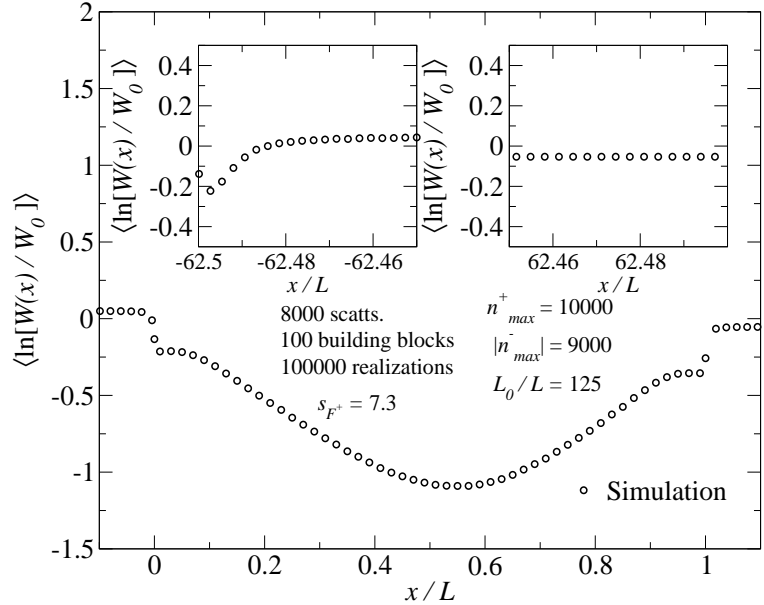}
}
\caption{
A computer simulation for the average of the logarithm of the electron density at zero temperature, when the left reservoir has a 
higher chemical  potential then the right one ($\mu_1 > \mu_2$).
It is expressed in units of its value for a non-disordered system, 
$\langle W(x)/W_0 \rangle$.
The insets show the density outside the sample in more detail.
The value $L_0/L= 125$ was used.
The ratio of the sample length to the mfp for the left Fermi level
is $s_{F^{+}}=L/\ell_F = 7.3$;
the ratio of the number of states between the lower and upper chemical potentials $\Delta N=n_{max}^{+}-\vert n_{max}^{-}\vert$ and the total number of states $N$
is $\Delta N/N=0.053$.
The details of the simulation are the same as those explained in Fig. \ref{density for L and for R incid and all levels 1}.
}
\label{log W(x) simul}
\end{figure}

Fig. \ref{log W(x) simul} shows the contribution of $N=19000$ states ($n_{max}^{+}=10000$, $\vert n_{max}^{-}\vert =9000$) to the average of the logarithm of the electron density inside and outside the system.
We only present a computer simulation, as we do not have a theoretical prediction for this case.

\section{Summary and Conclusions}
\label{conclusions}

In this paper we studied the electron density in a
problem of electronic transport in a
one-dimensional disordered multiply-scattering conductor:
we analyzed the contribution of the individual electron energy levels
and their total contribution.

A model was proposed for the density matrix of the system placed between two reservoirs at the same temperature $T$, but, in general, different chemical potentials $\mu_1, \mu_2$. 
The model is given in
Eq. (\ref{model density matrix disordered}), and depends on the temperature $T$ and the chemical potentials of the two reservoirs, $\mu_1, \mu_2$.
With its aid, the statistical-mechanical expectation value of the electron density was evaluated. 
The system is not in equilibrium, but is supposed to be in a stationary
state.

We then computed an average over an ensemble of configurations of disorder.
The theoretical analysis was performed within the DMPK model \cite{mello-kumar,mpk} and the results were compared  with computer simulations.
We studied the statistics of the {\it electron density} and of its 
{\it logarithm} over such an ensemble of configurations,
concentrating on the zero-temperature limit, $T=0$.

We first considered the situation in which the system is in equilibrium with 
the two reservoirs, i.e., $\mu_1=\mu_2$.

For individual energy levels way up in the spectrum, the DMPK predictions are generally very good for the average electron density, Fig \ref{density for L and for R incid c 1},
 and also for the average of its logarithm, 
Fig. \ref{aver ln W level n}.

Further down in the spectrum, the assumptions behind DMPK are not applicable.
However, the nature of the results is physically well understood.
As we go down in energy, the wave function penetrates ever less inside the sample.
The interpretation of $s_n$ in terms of a mfp due to localization from disorder gradually gives way to an interpretation as a parameter that measures the extent to which 
the wave function is reflected back because each scatterer is 'seen' by the electron as a higher and higher, and hence impenetrable, potential barrier.
This is reflected in the fact that the LDOS \cite{datta} is depleted in the interior of the
system, since the wave function is ever smaller inside.

The logarithm of the electron density is a self-averaging quantity: as a result, its average value computed with a finite but large sample shows very small statistical fluctuations.

As a consequence of the individual-level contributions, the DMPK prediction for the average of the total electron density,  i.e., the density summed over all the energy levels up to the Fermi energy, while being very good in the ballistic regions, 
is not adequate inside the disordered sample: again, the physical origin of the discrepancy is well understood.

The above results correspond to an equilibrium situation. Out of equilibrium, when the two chemical potentials are different, i.e., $\mu_1 \neq \mu_2$, the DMPK prediction
for the average of the total density is very good in the ballistic regions, while inside the sample it has a behavior similar to that of the above discussion. 
For the logarithm, we only present computer simulations.

We have not been able to find analytical results for other statistical quantities, like the variance of the electron density.
This, and other quantities, will have to wait for further analytical developments.

The equivalence between electronic conductance and the transmittance, allows to extend some predictions found in the present electronic study to classical systems; however, it is important to take into account the differences between quantum systems and classical ones.

Finally, we should remark that it would be desirable to find a 'gedanken' experiment designed to measure the electronic density, and compare it with our theoretical predictions: so far, we have not succeeded in this endeavor. 

\section*{Acknowledgements}

The authors thank F. Leyvraz, B. Shapiro and A. Z. Genack for their comments and suggestions.
The authors thank C. Lopez Nataren for technical support in the numerical simulations.
G. Rivas and M. Y\'epez thank Dr. Pier A. Mello, who unfortunately died during the review of this article: mentor, friend and a great human being. Rest in peace.

\section*{Funding}
GR is financially supported by the PhD scholarship of CONACyT, under Contract No. 777351.
MY and PAM are financially supported by the Sistema Nacional de Investigadores (SNI).
PAM was also supported by CONACyT, under Contract No. 282927.

\bibliographystyle{tfnlm}

\begin{thebibliography}{99}

\providecommand{\url}[1]{\normalfont{#1}}
\providecommand{\urlprefix}{Available from: }

\bibitem{mello-kumar}
Mello~PA, Kumar~N. Quantum transport in mesoscopic systems. complexity and
  statistical fluctuations. Oxford: Oxford University Press; 2010.

\bibitem{landauer}
Landauer~R. Electrical resistance of disordered one-dimensional lattices.
  Philosophical Magazine. 1970;\hspace{0pt}21:863--867.

\bibitem{buettiker}
B\"uttiker~M. Symmetry of electrical conduction. IBM Journal of Research and
  Development. 1988;\hspace{0pt}32:317--334.

\bibitem{Thouless_1977}
Thouless~DJ. Maximum metallic resistance in thin wires. Phys Rev Lett.
  1977;\hspace{0pt}39:1167--1169.

\bibitem{Abrahams_1979}
Abrahams~E, Anderson~PW, Licciardello~DC, et~al. Scaling theory of
  localization: Absence of quantum diffusion in two dimensions. Phys Rev Lett.
  1979;\hspace{0pt}42:673--676.

\bibitem{Anderson_1980}
Anderson~PW, Thouless~DJ, Abrahams~E, et~al. New method for a scaling theory of
  localization. Phys Rev B. 1980;\hspace{0pt}22:3519--3526.

\bibitem{Melnikov_1980}
Mel'nikov~VI. Distribution of resistivity probabilities of a finite, disordered
  system. Pis'ma Zh Ekps Teor Fiz. 1980;\hspace{0pt}32:244--247. [{\it JETP
  Lett}. 1980;32:225-228]; Fluctuations in the resistivity of a finite
  disordered system. Fis Tverd Tela (Leningrad). 1981;23:782–786. [{\it Sov.
  Phys. Solid State}. 1981;23:444-446].

\bibitem{Dorokhov_1984}
Dorokhov~ON. On the coexistence of localized and extended electronic states in
  the metallic phase. Solid State Commun. 1984;\hspace{0pt}51:381.

\bibitem{van_Rossum_1999}
van Rossum~MCW, Nieuwenhuizen~TM. Multiple scattering of classical waves:
  microscopy, mesoscopy, and diffusion. Rev Mod Phys. 1999;\hspace{0pt}71:313.

\bibitem{Shi_2014}
Shi~Z, Wang~J, Genack~AZ. Microwave conductance in random waveguides in the
  cross-over to Anderson localization and single-parameter scaling. Proceedings
  of the National Academy of Sciences. 2014;\hspace{0pt}111:2926--2930.

\bibitem{Dietz_2011}
Dietz~O, Kuhl~U, St\"ockmann~HJ, et~al. Microwave realization of
  quasi-one-dimensional systems with correlated disorder. Phys Rev B.
  2011;\hspace{0pt}83:134203.

\bibitem{dahr}
Dhar~A. Heat transport in low-dimensional systems. Advances in Physics.
  2008;\hspace{0pt}57:457--537.

\bibitem{gazaryan}
Gazaryan~YL. The one-dimensional problem of propagation of waves in a medium
  with random inhomogeneities. Sov Phys JETP. 1969;\hspace{0pt}29:996--1003.

\bibitem{kohler_papanicolau}
Kohler~W, Papanicolau~GC. Power statistics for wave propagation in one
  dimension and comparison with radiative transport theory. Journal of
  Mathematical Physics. 1973;\hspace{0pt}14:1973.
  
\bibitem{genack1990long}
Genack~A, Garcia~N, Polkosnik~W. Long-range intensity correlation in random media. Physical Review Letters. 1990;\hspace{0pt}65(17):2129.
  
\bibitem{Neupane-Yamilov}
Neupane~P, Yamilov~AG. Applicability of the position-dependent diffusion
  approach to localized transport through disordered waveguides. Phys Rev B.
  2015;\hspace{0pt}92:014207.

\bibitem{van_tiggelen_et_al_2000-2006}
van Tiggelen~BA, Lagendijk~A, Wiersma~DS. Reflection and transmission of waves
  near the localization threshold. Phys Rev Lett. 2000;\hspace{0pt}84:4333.

\bibitem{van_tiggelen_et_al_2000-2006_bis}
Skipetrov~SE, van Tiggelen~BA. Dynamics of Anderson localization in open 3d
  media. Phys Rev Lett. 2006;\hspace{0pt}96:043902.

\bibitem{tian_et_al_2010}
Tian~CS, Cheung~SK, Zhang~ZQ. Local diffusion theory for localized waves in
  open media. Phys Rev Lett. 2010;\hspace{0pt}105:263905.

\bibitem{mello-shi-genack}
Mello~PA, Shi~Z, Genack~AZ. Connection between wave transport through
  disordered 1d waveguides and energy density inside the sample: A
  maximum-entropy approach. Physica E: Low-dimensional Systems and
  Nanostructures. 2015;\hspace{0pt}74:603 -- 607.

\bibitem{cheng-yepez-mello-genack}
Cheng~X, Ma~X, Y\'epez~M, et~al. Single-parameter scaling and maximum entropy
  inside disordered one-dimensional systems: Theory and experiment. Phys Rev B.
  2017;\hspace{0pt}96:180203(R).

\bibitem{boris_1986}
Shapiro~B. Large intensity fluctuations for wave propagation in random media.
  Phys Rev Lett. 1986;\hspace{0pt}57:2168--2171.

\bibitem{boris_2014}
Sarma~R, Yamilov~A, Neupane~P, Shapiro~B, and Cao~H.
Probing long-range intensity correlations inside disordered photonic nanostructures.
Physical Review B 2014;\hspace{0pt}90(1):014203.
 
\bibitem{bender_coherent_2022}
Bender~N, Goetschy~A, Hsu~CW, Y{\i}lmaz~H, Palacios~PJ, Yamilov~A, and Cao~H.
Coherent enhancement of optical remission in diffusive media. 
Proceedings of the National Academy of Sciences. 2022;\hspace{0pt}119(41):e2207089119.

\bibitem{bender_depth-targeted_2022}
Bender~N, Yamilov~A, Goetschy~A, Y{\i}lmaz~H, Hsu~CW, and Cao~H. Depth-targeted energy delivery deep inside scattering media. Nat. Phys. 2022;\hspace{0pt}18(3):309–315.

\bibitem{bender_introducing_2019}
Bender~N, Y{\i}lmaz~H, Bromberg~Y, and Cao~H.
Introducing non-local correlations into laser speckles.
Optics Express 2019;\hspace{0pt}27(5):6057--6067.

\bibitem{koirala_inverse_2019}
Koirala~M, Sarma~R, Cao~H, and Yamilov~A.
Inverse design of long-range intensity correlation in scattering media.
Physical Review B 2019;\hspace{0pt}100(6):064203.

\bibitem{sarma_control_2015}
Sarma~R, Yamilov~A, Liew~SF, Guy~M, and Cao~H.
Control of mesoscopic transport by modifying transmission channels in opaque media.
Physical Review B 2015;\hspace{0pt}92(21):214206.

\bibitem{sarma_using_2015}
Sarma~R, Yamilov~A, Neupane~P, and Cao~H.
Using geometry to manipulate long-range correlation of light inside disordered media.
Physical Review B 2015;\hspace{0pt}92(18):180203.

\bibitem{sarma_control_2016}
Sarma~R, Yamilov~AG, Petrenko~S, Bromberg~Y, and Cao~H.
Control of energy density inside a disordered medium by coupling to open or closed channels.
Physical Review Letters 2016;\hspace{0pt}117(8):086803.

\bibitem{yamilov_shape_2016}
Yamilov~A, Petrenko~S, Sarma~R, and Cao~H.
Shape dependence of transmission, reflection, and absorption eigenvalue densities in disordered waveguides with dissipation.
Physical Review B 2016;\hspace{0pt}93(10):100201.

\bibitem{yamilov_sum_2022}
Yamilov~A, Bender~N, and Cao~H.
Sum rules for energy deposition eigenchannels in scattering systems.
Optics Letters 2022;\hspace{0pt}47(19):4889--4892.

\bibitem{yilmaz_2019_transverse}
Y{\i}lmaz~H, Hsu~CW, Yamilov~A, and Cao~H.
Transverse localization of transmission eigenchannels.
Nature Photonics 2019;\hspace{0pt}13(5):352--358.

\bibitem{elastic_waves}
Calleja~\'Angel~J, Torres~Guzm\'an~JC, D\'iaz~de Anda~A. Anderson localization
  of flexural waves in disordered elastic beams. Scientific Reports.
  2019;\hspace{0pt}9:3572.

\bibitem{elastic_waves_2}
Flores-Olmedo~E, Mart\'inez-Arg\"uello~AM, Mart\'inez-Mares~M, et~al.
  Experimental evidence of coherent transport. Scientific Reports.
  2016;\hspace{0pt}6:25157.

\bibitem{Erdos}
Erd\"os~P, Herndon~RC. Theories of electrons in one-dimensional disordered
  systems. Advances in Physics. 1982;\hspace{0pt}31:65--163.

\bibitem{Shapiro_1986}
Shapiro~B. Probability distributions in the scaling theory of localization.
  Phys Rev B. 1986;\hspace{0pt}34:4394--4397.

\bibitem{mello_1987}
Mello~PA. Macroscopic approach to the theory of one-dimensional disordered
  conductors. Phys Rev B. 1987;\hspace{0pt}35:1082--1087.

\bibitem{Aspect_2009}
Aspect~A, Inguscio~M. Anderson localization of ultracold atoms. Physics Today.
  2009;\hspace{0pt}62:30--35.

\bibitem{Jean-Philippe}
Brantut~JP, Meineke~J, Stadler~D, et~al. Conduction of ultracold fermions
  through a mesoscopic channel. Science. 2012;\hspace{0pt}337:1069--1071.

\bibitem{Billy_et_al}
Billy~J, Josse~V, Zhanchun~Z, et~al. Direct observation of Anderson
  localization of matter waves in a controlled disorder. Nature.
  2008;\hspace{0pt}453:891--894.

\bibitem{Roati_et_al}
Roati~G, D’Errico~C, Fallani~L, et~al. Anderson localization of a
  non-interacting bose-einstein condensate. Nature.
  2008;\hspace{0pt}453:895--898.

\bibitem{Huang_2022}
Huang~Y, Kang~Y, Genack~AZ. Wave excitation and dynamics in non-hermitian
  disordered systems. Phys Rev Research. 2022;\hspace{0pt}4:013102.

\bibitem{Carminati}
Carminati~R, Chen~H, Pierrat~R, et~al. Universal statistics of waves in a
  random time-varying medium. Phys Rev Lett. 2021;\hspace{0pt}127:094101.

\bibitem{buttiker_larmor_nodate}
B\"uttiker~M. Larmor precession and the traversal time for tunneling.
Phys. Rev. B. 1983;\hspace{0pt}27:6178--6188.

\bibitem{winful_delay_2003}
Winful~HG. Delay time and the Hartman effect in quantum tunneling.
Phys. Rev. Lett. 2003;\hspace{0pt}91:260401.

\bibitem{iannaccone_general_1994}
Iannaccone~G. General relation between density of states and dwell times in mesoscopic systems.
Phys. Rev. B. 1995;\hspace{0pt}51:4727--4729.

\bibitem{shi2015dynamic}
Shi~Z, Genack~AZ. 
Dynamic and spectral properties of transmission eigenchannels in random media.
Physical Review B. 2015;\hspace{0pt}92(18):184202.

\bibitem{choi}
Choi~W, Mosk~AP, Park~QH, et~al.
Transmission eigenchannels in a disordered medium.
Phys. Rev. B. 2011;\hspace{0pt}83:134207.

\bibitem{davy2015transmission}
Davy~M, Shi~Z, Wang~J, et~al.
Transmission eigenchannels and the densities of states of random media.
Physical Review Letters 2015;\hspace{0pt}114(3):033901.

\bibitem{bender2020fluctuations}
Bender~N, Yamilov~A, Y{\i}lmaz~H, et~al.
Fluctuations and correlations of transmission eigenchannels in diffusive media.
Physical Review Letters. 2020;\hspace{0pt}125(16):165901.

\bibitem{mello-yepez}
Mello~PA, Y\'epez~M. Electron transport and electron density inside
  quasi-one-dimensional disordered conductors. Phys Rev B.
  2020;\hspace{0pt}101:014206.

\bibitem{mpk}
Mello~PA, Pereyra~P, Kumar~N. Macroscopic approach to multichannel disorder
  conductors. Ann Phys (NY). 1988;\hspace{0pt}181:290--317.

\bibitem{mello-shapiro-imry}
Mello~PA, Imry~Y, Shapiro~B. Model for phase breaking in the electronic
  conduction in mesoscopic systems. Phys Rev B.
  2000;\hspace{0pt}61:16570--16581.

\bibitem{dhar_saito_hanggi}
Dhar~A, Saito~K, H\"anggi~P. Nonequilibrium density-matrix description of
  steady-state quantum transport. Phys Rev E. 2012;\hspace{0pt}85:011126.

\bibitem{froufe_et_al_pre_2007}
Froufe-P\'erez~LS, Y\'epez~M, Mello~PA, et~al. Statistical scattering of waves
  in disordered waveguides: From microscopic potentials to limiting macroscopic
  statistics. Phys Rev E. 2007;\hspace{0pt}75:031113.

\bibitem{yepez-saenz}
Y\'epez~M, S\'aenz~JJ. Contribution of evanescent waves to the effective medium
  of disordered waveguides. EPL (Europhysics Letters).
  2014;\hspace{0pt}108:17006.
  
\bibitem{Beenakker_1997}
Beenakker~CWJ.
Random-matrix theory of quantum transport.
Reviews of Modern Physics. 1997;\hspace{0pt}69:731.

\bibitem{datta}
Datta~S. Electronic transport in mesoscopic systems. Cambridge: Cambridge
  University Press; 1995.

\bibitem{economou2006green}
Economou~EN. Green's functions in quantum physics. Vol.~7. Springer Science \&
  Business Media; 2006.

\end{thebibliography}

\appendix

\section{The model for the density matrix of Ref.  \cite{dhar_saito_hanggi}}
\label{dhar_et_al_density_matrix}

In  Ref.  \cite{dhar_saito_hanggi}, the authors study a tight-binding approach of non-interacting electrons,
there being $N_{sites}$ sites 
for the system proper. 
They consider the correlation matrix 
$\langle  c^{\dagger}_m c_l\rangle$, with $m, l = 1, \cdots , N_{sites}$ numbering the sites, 
and designate by $d_s, \;  s=1,\cdots N$, the eigenvalues of this matrix.

In Eqs. (2.11)-(2.14) of Ref.  \cite{dhar_saito_hanggi}, the system reduced density matrix in the stationary state is found to be
\begin{subequations}
\begin{eqnarray}
\hat{\rho_S}
= \prod_{n=1}^{N_{sites}}\frac{ e^{-a_n {c'_n}^{\dagger}{c'_n}} }{1+e^{-a_n}} \; ,\\
a_n = \ln \left(\frac{1}{d_n} - 1 \right).
\end{eqnarray}
\label{dhar reduced_rho_S}
\end{subequations}
In  Sec. IIIA of Ref.  \cite{dhar_saito_hanggi} it is found, in the {\it weak-coupling limit},
\begin{subequations}
\begin{eqnarray}
a_n &=& \ln \left(\frac{1}{e_n} - 1 \right) \; ,
\end{eqnarray}
where
\begin{eqnarray}
e_n &=& \gamma_L^n \; f(\epsilon_n, \mu_L, T_L)
 + \gamma_R^n \; f(\epsilon_n, \mu_R, T_R), 
\end{eqnarray}
with
\begin{eqnarray}
\gamma_L^n + \gamma_R^n &=& 1,
\end{eqnarray}
where 
\begin{eqnarray}
f(\epsilon_n, \mu, T) = \frac{1}{1+ e^{\beta(\epsilon_n - \mu)}} \; .
\label{fermi}
\end{eqnarray}
\label{dhar as es}
\end{subequations}
denotes the Fermi function.

In the present main text, $n$ designates 
'running-wave states', with 
$n=0, \pm1, \pm2, \cdots$,
instead of the `site states' of Ref.  \cite{dhar_saito_hanggi}.
The model of Eq. (\ref{model density matrix}) in the text is obtained by setting, in the above equations,
$T_L=T_R=T$, $\mu_L=\mu_1, \mu_R=\mu_2$, and identifying
\begin{subequations}
\begin{eqnarray}
&& \gamma_L^{n>0} = 1, \;\;\; \gamma_R^{n>0} = 0 \;\;\; 
\Rightarrow \;\;\; 
a_{n>0} = \ln\left[ \frac{1}{f(\epsilon_n, \mu_L, T)} - 1\right]
= \beta(\epsilon_n - \mu_L)  
\label{s>0}  \\
&& \gamma_L^{n=0} = \gamma_R^{n=0} = \frac12    \;\;\; 
\Rightarrow \;\;\; 
e_{n=0}=\frac12 \left[f(\epsilon_n=0, \mu_L, T) +  f(\epsilon_n=0, \mu_R, T)  \right] 
\nonumber    \\
&& \hspace{5.4cm} \equiv f(\epsilon_n=0, \mu_0, T)  
 \;\;\; \Rightarrow \;\;\; 
a_{n=0}= -\beta \mu_0     
\label{s=0}   \\
&& 
\gamma_L^{n<0} =0, \;\;\; \gamma_R^{n<0} = 1
\;\;\; \Rightarrow \;\;\; 
a_{n<0} = \ln\left[ \frac{1}{f(\epsilon_n, \mu_R, T)} - 1\right]
= \beta(\epsilon_n - \mu_R)
\label{s<0}
\end{eqnarray}
\end{subequations}
in the above equations.
In Eq. (\ref{s=0}), $\mu_0 \in (\mu_L, \mu_R)$.

\section{Proof of Eq. (\ref{W(x) inside})}
\label{proof eq Wn+}

We have denoted by $\alpha_i, \beta_i$, $i=1,2$ the elements of the transfer matrix $M_i$ 
for the portions 1 and 2 of the sample, on the left and right of the observation point $x$, respectively, i.e.,
\begin{equation}
M_i = 
\left[
\begin{array}{cc}
\alpha_i   & \beta_i \\
\beta_i^{*} & \alpha_i^{*}
\end{array}
\right] , \;\;\;\;\; i=1,2,
\label{M_i}
\end{equation}
with the condition $|\alpha_i|^2 - |\beta_i|^2 =1$, thus satisfying the requirements of time-reversal invariance and flux conservation.
When no index $i$ is employed, we shall understand the various quantities to refer to the wire as a whole.

Let $a$ and $b$ denote the amplitudes of the right-going and left-going waves at the point $x$, and $0$ and $t$ 
($t$=transmission amplitude) the amplitudes outside the wire on the right-hand side.
Then, from the definition of the transfer matrix $M_2$ we have
\begin{equation}
M_2
\left[
\begin{array}{c}
a    \\
b 
\end{array}
\right]
=
\left[
\begin{array}{c}
t   \\
0 
\end{array}
\right] \; .
\label{a,b,t 1}  
\end{equation}
We invert this equation to find $a$ and $b$, making use of the relation
\begin{equation}
M_2^{-1}
=\left[
\begin{array}{rr}
\alpha_2^{*}   & -\beta_2 \\
-\beta_2^{*} & \alpha_2
\end{array}
\right] \; ,
\label{M_2-inv}
\end{equation}
to find
\begin{equation}
\left[
\begin{array}{c}
a    \\
b 
\end{array}
\right]
=
\left[
\begin{array}{r}
t \alpha_2^{*}  \\
-t \beta_2^{*}
\end{array}
\right] .
\label{a,b,t 2}  
\end{equation}
We thus have
\begin{eqnarray}
|a e^{ik_{n}x} + be^{-ik_{n}x}|^2  
&=& T|\alpha_2^{*}e^{ik_{n}x} - \beta_2^{*}e^{-ik_{n}x}|^2
\equiv T F_{n^+}(M_2) ,
\label{I(z) b} 
\end{eqnarray}
which gives the result appearing in the first line of Eq. (\ref{W(x) inside}).

Similarly, from the definition of the transfer matrix $M_1$, we have
\begin{equation}
\left[
\begin{array}{c}
a'  \\
b'
\end{array}
\right]
=M_1
\left[
\begin{array}{c}
0    \\
t' 
\end{array}
\right]
=\left[
\begin{array}{rr}
\alpha_1   & \beta_1 \\
\beta_1^{*} & \alpha_1^{*}
\end{array}
\right] 
\left[
\begin{array}{c}
0    \\
t' 
\end{array}
\right]
=\left[
\begin{array}{c}
\beta_1 t'   \\
\alpha_1^{*}t' 
\end{array}
\right]  .
\label{a,b,t 1}  
\end{equation}
We thus have
\begin{subequations}
\begin{eqnarray}
\left| a' e^{ik_{n}x} + b'e^{-ik_{n}x} \right|^2  
&=& \left|\beta_1e^{ik_{n}x} + \alpha_1^{*}e^{-ik_{n}x}\right|^2 T'
\equiv T F_{n^-}(M_1) ,
\label{I(z) b} 
\label{I(z) c}
\end{eqnarray}
\label{I(z)}
\end{subequations}
which gives the result appearing in the second line of Eq. (\ref{W(x) inside}).

\section{Melnikov's equation for $p_s(\lambda)$ and first and second moments of 
$\lambda$}
\label{1st and 2nd mom of lambda}

The probability density $p_s(\lambda)$ satisfies Melnikov's equation \cite{Melnikov_1980}
--which is the particular case of the DMPK equation for one open channel-- 
given by 
[see also  Ref.  \cite{mello_1987}]
\begin{equation}
\frac{\partial p_s(\lambda)}{\partial s}
= \frac{\partial}{\partial \lambda}
\left[
\lambda(1+ \lambda) \frac{\partial  p_s(\lambda)}{\partial \lambda}
\right].
\label{melnikov}
\end{equation}

The first and second moments of $\lambda$ are given by
\begin{subequations}
\begin{eqnarray}
\langle \lambda \rangle_s 
&=& \frac12 \left( e^{2s} -1 \right), \\
\langle \lambda^2 \rangle_s 
&=& \frac{1}{12}\left(
2 e^{6s}- 6 e^{2s} + 4
\right).
\end{eqnarray}
\end{subequations}

\section{
Local density of states and the dwell time}
\label{DOS and Tau}

Consider the one-dimensional electronic system of Fig. \ref{1D and two reservoirs}. For a given energy level $\epsilon_{n}$, the local density of states (LDOS) at a point $x$ inside the disordered system is the sum of the particle densities given rise from the left and right incidences \cite{economou2006green}, i.e., 
\begin{equation}
\bm\rho_{_{LDOS}}\left(\epsilon_{n},x\right)=\frac{L_{0}}{2\pi \hslash}
\frac{1}{v_{n}}
\left[ 
\vert \psi_{n}^{L_{0}} \left(x\right)  \vert ^{2} +
\vert \psi_{-n}^{L_{0}} \left(x\right) \vert ^{2}
\right],
\;\; v_{n}=\frac{\hslash k_{n}}{m}=\sqrt{\frac{2\epsilon_{n}}{m}}.
\label{Def_LDOS_1}
\end{equation}
here, $v_{n}$ in the incident flux, which in 1D is the group velocity, Eq. \eqref{density of states c}. The wave functions $\psi^{L_0}_{n}\left( x\right) $ are taken from Table \ref{structure of wf 2} and we have assumed $n>0$.

Since the dimensionless density for the $n$-th level $w_{n}\left(x\right)$, Eq. \eqref{Individual_level_w_n}, is the sum of the $n$-th level contributions for left incidence $ w_{n}^{LI} \left(x\right)$ and right incidence $w_{n}^{RI} \left(x\right)$, Eq. \eqref{wn_LI_RI}, then the LDOS can be written as
\begin{equation}
\bm\rho_{_{LDOS}}\left(\epsilon_{n},x\right)=\frac{L_{0}}{2\pi \hslash}
\frac{1}{v_{n}}\frac{1}{L_{0}}
\left[   w_{n}^{LI} \left(x\right) + w_{n}^{RI} \left(x\right)  \right]
=
\frac{1}{2\pi \hslash} \frac{\left[ 2w_{n}\left(x\right) \right]}{v_{n}};
\label{Def_LDOS_2}
\end{equation}
therefore, the dimensionless density is related to the LDOS as
\begin{equation}
w_{n}\left(x\right)
= \frac{ v_{n} }{2}
\left[ 2\pi \hslash \bm\rho_{_{LDOS}}\left(\epsilon_{n},x\right) \right];
\label{Def_LDOS_3}
\end{equation}

From the LDOS, Eq. \eqref{Def_LDOS_1}, the Density of States (DOS) inside the disordered sample is obtained in the following way:
\begin{equation}
\bm\rho_{_{DOS}}\left(\epsilon_{n}\right)
=\int_{0}^{L} \bm\rho_{_{LDOS}}\left(\epsilon_{n},x\right) dx.
\label{total DOS} 
\end{equation}

The DOS is related to the time spent by the particle inside the disordered sample before being reflected or transmitted; this characteristic time is called dwell time $\tau_{_{D}}\left( \epsilon_{n}\right)$, which is proportional to the number of particles inside the sample and inversely proportional to the incidence flux, i.e.,
\begin{equation}
\tau_{_{D}}  \left( \epsilon_{n}\right)
=
\frac{L_{0}}{v_{n}}
\int_{0}^{L}
\left[ 
\vert \psi_{n}^{L_{0}} \left(x\right)  \vert ^{2} +
\vert \psi_{-n}^{L_{0}} \left(x\right) \vert ^{2}
\right]dx,
\label{dwell time}
\end{equation}
or in terms of the dimensionless electron densities $w_{n}^{LI}\left( x\right)$ and $w_{n}^{RI}\left( x \right)$,
\begin{equation}
\tau_{_{D}}  \left( \epsilon_{n}\right)
=
\frac{1}{v_{n}}
\int_{0}^{L}
\left[ w_{n}^{LI} \left(x\right) + w_{n}^{RI} \left(x\right) \right]dx
=
\frac{1}{v_{n}}
\int_{0}^{L} 2 w_{n} \left(x\right)dx.
\end{equation}
therefore
\begin{equation}
\tau_{_{D}}  \left( \epsilon_{n}\right)
=2\pi \hslash \bm\rho_{_{DOS}}\left(\epsilon_{n}\right)
\end{equation}

The general relation between the dwell time and the DOS in mesoscopic media is found in Ref. \cite{iannaccone_general_1994}.


\end{document}